\documentclass[12pt,preprint]{aastex}

\shorttitle{Energetic emissions from the M87~jet}
\shortauthors{Honda \& Honda}

\begin{document}

\title{TRANSITIVE X-RAY SPECTRUM AND PeV GAMMA-RAY CUTOFF
IN THE M87~JET: ELECTRON "PEVATRON"}
\author{Mitsuru~Honda}
\affil{Plasma Astrophysics Laboratory, Institute for Global Science,
Mie, Japan}
\author{Yasuko~S.~Honda}
\affil{Department of Electrical and Information Engineering,\\*
Kinki University Technical College, Mie, Japan;\\*
yasuko@ktc.ac.jp}

\begin{abstract}
We propose a modified version of the X-ray spectral index and an
intrinsic cutoff frequency of inverse Compton radiation from
the brightest knot of the M87 jet, in conjunction with an
application of the new conceptions of injection and
diffusive shock acceleration (DSA) of electrons in
magnetized filamentary plasma to the specified source.
The drop of the X-ray flux density in a transitive frequency
region is associated with the interplay of ordinary
synchrotron cooling and weaker magnetic fields
concomitant with the smaller scale filaments that
allow the electron injection, while the radio-optical
synchrotron continuum is dominantly established by the
major electrons that are quasi-secularly bound to larger filaments.
With reference to, particularly, the updated external Compton model,
we demonstrate that in the Klein-Nishina regime fading inverse
Comptonization, the injected electrons can be stochastically
energized up to a Lorentz factor as high as $5\times 10^{10}$
in the temporal competition with diffuse synchrotron cooling;
this value is larger than that attainable for a simple
DSA scenario based on the resonant scattering diffusion
of the gyrating electrons bound to a supposed magnetic
field homogeneously pervading the entire knot.
The upper limits of the photon frequency boosted via
conceivable inverse Compton processes are predicted to
be of the common order of $\sim 10^{30}~{\rm Hz}$.
The variability of the broadband spectrum is also discussed
in comparison to the features of a blazar light curve.
The present scenario of a peta-eV (PeV; $10^{15}~{\rm eV}$)
electron accelerator, the "Pevatron," might provide some
guidance for exploring untrod hard X-ray and gamma-ray
bands in forthcoming observations.
\end{abstract}


\section{INTRODUCTION}

A fundamental question --- the observable highest frequency of
radiation from extragalactic jets --- remains unresolved, as the
closely relevant physics of the particle acceleration incorporated
with injection and the cosmological transport of energetic photons
is largely unknown in both theoretical and observational aspects.
As a candidate celestial object to study those issues in detail,
of a suitable nature is a nearby powerful radio source,
particularly the M87~jet, and so far, many efforts have been
devoted to measure the photon spectra in higher frequency
bands (optical, \citealt{meisenheimer96}; \citealt{perlman01};
X-ray, \citealt{marshall02}; \citealt{wilson02}).
In the previous key work that was based on ground-based
observations \citep{meisenheimer96}, the synchrotron spectra
were suggested to, in the entire jet, have an almost
constant cutoff of $(0.8-2.5)\times 10^{15}~{\rm Hz}$
\citep[see also][]{perez-fournon88,stiavelli97}, compared
to a universal value among radio jets \citep{keel88}.
The corresponding Lorentz factor of the electrons was
estimated to be $9\times 10^{5}$, which could be deduced
from a simple model of the diffusive shock acceleration (DSA)
that relied on the small-angle resonant scattering by the
turbulent Alfv\'en waves \citep{biermann87,longair92}.
However, state-of-the-art {\it Chandra} observations revealed that
the spectra indicate no steep cutoffs, extending to X-ray regions
\citep{marshall02}, far beyond the previous cutoff.
The X-ray continua have been ascribed to synchrotron
radiation, whereas inverse Compton (IC) models as a
dominant X-ray mechanism have been ruled out.
The Lorentz factor of the X-ray--emitting electrons is
considered to be of order $\sim 10^{6}-10^{8}$
\citep{meisenheimer96,perlman01,marshall02,wilson02},
although the achievable maximum is as yet poorly known.

A clue to this puzzling question is that simple models invoking shocks
and synchrotron losses in a uniformly filled jet overpredict the
X-ray flux density by large factors \citep{biretta91,marshall02}.
In fact, the fine structure of magnetized filaments including
the limb-brightening feature has been found at many positions
along the jet \citep{reid89,owen89,stiavelli91,lobanov03}.
The cancellation of the small-scale magnetic fields could be
responsible for the observational fact that in the interior of
the jet, the magnetic field is significantly free \citep{owen89},
reducing the polarization \citep{capetti97}.
Accordingly, the requirement for modeling a nonuniform jet is
that in the interior the turbulent state ought to be rather vital,
in contrast to conventional jet models involving a turbulent
radial edge \citep[e.g.,][]{ostrowski00} caused by surface-mode
instabilities such as the Kelvin-Helmholtz instability.

At this juncture, the novel model by \citet{honda02}, in which
the jet is hypothesized to comprise the magnetized filaments
that are self-organized via a current filamentation instability
\citep{honda00a}, seems to satisfy this requirement.
In their model, the vectors of fluctuating magnetic fields in the
interior are supposed to be randomly oriented in the transverse
direction of the filaments, such that they could be macroscopically
canceled out, although in the envelope region, the uncanceled,
ordered magnetic field could appear \citep{honda04a,honda05a}.
By invoking a similar environment, \citet{fleishman06} has
recently examined synchrotron radiation from the electrons
injected into the small-scale random magnetic fields and found
that the diffuse emission including higher harmonics reproduced
the observational X-ray continuum at the knots~B and C1.
However, the mechanisms of the electron injection reflecting
the energy hierarchy that ought to be involved in the filamentary
plasma \citep{honda05a} and the related in situ acceleration
were not self-consistently taken into consideration.
It is important to investigate these details, because the complex
interplay will actually connect the confirmed radio-optical
continuum with the diffuse synchrotron spectrum in the energetic
regime to give a transitive spectrum compared to the observed
X-ray spectrum \citep{marshall02}, which is expected to deviate
from the conventional model spectrum simply involving the
synchrotron cooling \citep[e.g.,][]{kardashev62,longair94}.
Also, the injected electrons quickly lose the coherency of
their trajectory in the filamentary (stochastic) medium
and thereby have a shorter mean free path (mfp) that deviates
from the universal gyroscaling (valid for the simple
case in which a mean magnetic field can be well defined);
this is just the case that is advantageous to the DSA.
The possible reboost of electron energy will, in turn,
upshift the IC gamma-ray cutoff frequency.

In the context of the filamentary jet model, here we propose a
novel scheme to construct the steepened X-ray synchrotron spectrum
(\S~2) and attempt to provide genuine uppermost energies of
an accelerated electron and photon at knot\,A (\S~3), where
X-ray flux is so prominent as to be comparable to
that from the galactic core \citep{marshall02,perlman05}.
The theory is based on the DSA model incorporated
with the new type of injection reflecting electron
escape from small-scale magnetic traps, which is
referred to as the ``transition injection'' (\S~2.1).
We obtain a modified X-ray spectral index consistent
with the large flux density reduction measured in the
{\it Chandra} regime and a nonlinearly evolving feature
of the current filamentation instability (\S~2.2).
In the conceivable radiation fields, including external Compton (EC)
targets \citep{stawarz03}, we find that the injected electrons
can be, in situ, accelerated up to the $10~{\rm PeV}$ range (\S~3.1).
It is shown that the flux density of IC radiation surpasses,
above the soft gamma-ray region, that of synchrotron radiation,
and the highest frequency reaches the theoretical upper limit
reflecting the maximum electron energy (\S~3.2).
The IC spectrum can be accommodated with the recent
High Energy Stereoscopic System (HESS)
observation of sub-TeV gamma rays emanating from the
galaxy region \citep{beilicke05}, although the origin
of these gamma rays is still unresolved.
The energy budget of the diffuse synchrotron radiation in
the gamma-ray spectral tail is also estimated (\S~3.3).
Moreover, we include a discussion concerning the
variability of the broadband spectrum (\S~4).
The presence of fine filaments turns out to be essential
for the transition injection, raising the flux
density level of the diffuse radiation and
engendering the high-frequency variation, and therefore the
observability of such filaments is also discussed (\S~5).

\section{SYNCHROTRON SPECTRUM EXTENDING TO THE X-RAY BAND
\\* WITHOUT STEEP CUTOFF}

Highlighting the conception of the transition injection
that mediates the energy hierarchy involved in a magnetized
filamentary plasma, we consider the electron preacceleration
below the injection energy.
First we recover the radio-optical synchrotron continuum
in the cold limit (\S~2.1), and second, we construct the
transitive X-ray spectrum (\S~2.2).

\subsection{\it Impacts of the Small Spatioscale Entailed in
the Transition Injection}

Begin by considering a low-energy state in which cosmic-ray
(test) electrons are bound to the magnetized current filaments with
various transverse sizes $\lambda$, exhibiting gyromotion.
Apparently, this can be compared to the low-$\beta$ plasma state,
and along with this, it seems to be reasonable to suppose
that the locally ordered fields, with a strength of
$|{\bf B}({\bf r})|$, permeating through the filaments,
are preferentially disturbed by Alfv\'en waves.
As is accommodated by the test particle approximation,
we presume a small-amplitude disturbance that does not
affect the zeroth-order magnetic fields sustained by the
bulk filaments (the validity is confirmed later).
In this aspect, recalling the consensus that the jet knots are
associated with shocks, the {\it gyrating} cold electrons can
presumably be accelerated by a conventional Fermi-based DSA
mechanism relying on the weak Alfv\'enic fluctuations.
For a possible case of the scattering center compression ratio
of $r\sim 4$ \citep[reflecting a high Mach number shock;][]
{vainio99,schlickeiser02}, the predicted synchrotron emission
from the nonthermal electrons is certainly compatible with
the observed radio-optical continuum of the power-law
flux density distribution ($F_{\nu}\propto\nu^{-\alpha}$),
which has indices around $\alpha\sim 0.5$ \citep{biretta91};
that is, assigning the $\alpha$-value to the canonical expression
$\alpha_{1}=(p-1)/2$ \citep[e.g.,][]{longair94} gives a
differential spectral index of the electron density of
$p=(r+2)/(r-1)\sim 2$ (for $dn/d\epsilon=\kappa\epsilon^{-p}$,
where $\epsilon=\gamma mc^{2}$).
It is precisely noted that in general, the relation of $r$
to the gas compression ratio depends on the plasma conditions,
such as the plasma $\beta$-value \citep[see][for the
detailed calculations]{vainio99,schlickeiser02}.
As long as the gyroradius of the accelerated electrons $r_{g,e}$
is smaller than the transverse size of the local filament trapping
them ($\sim\lambda$), such a DSA mechanism will safely operate.
With regard to energy restrictions, the acceleration of the
electrons bound to a large filament ought to be knocked down
by radiative losses before the escape from the filament
(temporal limit), whereas the electrons bound to a small
filament could escape via the energization (spatial limit).
It is, thus, anticipated that there exists a critical scale of filament,
$\lambda_{c}$, that discriminates between the large and small scales.
Furthermore, with reference to the consequence of kinetic simulations
that the amplification of magnetic fields accompanies
the coalescence of filaments accumulating larger filaments
\citep{honda00a,honda00b}, we consider a situation
in which the filaments with larger (smaller) values of $\lambda$
involve the larger (smaller) values of $|{\bf B}|$.
Here we invoke the scaling of
$|{\bf B}|\propto\lambda^{(\beta-1)/2}$, where $\beta$ corresponds to
the power-law spectral index of the magnetized filamentary
turbulence \citep[possibly larger than the spectral index
for the Kolmogorov turbulence;][]{montgomery79,honda05a}.

In order to quantitatively examine the $\lambda$ dependence
of the maximum Lorentz factor of an accelerated electron
$\gamma^{\ast}$, of practical use is an extended scheme
that can treat the oblique mean magnetic field
with respect to the shock normal direction (for the
methodological details, see \citealt{honda04b}).
We consider a simple model in which the energy density ratio
of the fluctuating magnetic field to $|{\bf B}|^{2}/(8\pi)$,
denoted as $b$ ($\ll 1$), is spatially constant, and the
scattering of electrons occurs when the wave-particle
resonance condition of $r_{g,e}\sim k^{-1}$ is satisfied, where
the wavenumber $k$ is supposed to be in the inertial range
of the power-law fluctuation spectrum with an index of
$\beta^{\prime}$ \citep[e.g., $\beta^{\prime}=5/3$
for the Kolmogorov-type fluctuations;][]{biermann87,honda04b}.
Note that the local minimum of $k$ (i.e., the correlation
length of the Alfv\'enic fluctuations) is compared to the
filament size $\lambda$, whose maximum is described as
${\tilde l}D$, where ${\tilde l}$ ($\lesssim 1$) and
$D$ denote the reduction factor of the filament correlation
length and the diameter of the jet, respectively \citep[typically,
${\tilde l}D=110~{\rm pc}$ is expected at knot\,A;][]{biretta91}.

As concerns the radiative losses of accelerated electrons,
we take normal synchrotron and IC cooling into account.
The former is ascribed to the gyration in the local mean field,
whose intensity can be expressed in the phenomenological form
of $|{\bf B}|=B_{m}(\lambda/{\tilde l}D)^{(\beta-1)/2}$,
where $B_{m}\equiv |{\bf B}|_{\lambda={\tilde l}D}$.
The diffuse effects considered in \S~3 are ignored because $b\ll 1$.
For the latter, it should be remarked that while the synchrotron
self-Compton (SSC) process can be preferentially ignored (\S~3.1),
influential EC candidates do exist, including the starlight
and extended dust emission of the host galaxy \citep{stawarz03}.
At knot\,A, which is at a distance of 1 kpc from the galactic core
\citep{honda04a}, the energy densities of the target photons are,
in the comoving (shock rest) frame, suggested to be of the levels
of $u_{\rm ph}\sim 10^{-9}\Gamma^{2}~{\rm ergs~cm^{-3}}$ (stars)
and $\sim 10^{-11}\Gamma^{2}~{\rm ergs~cm^{-3}}$ (dust), where
$\Gamma=(1-U^{2}/c^{2})^{-1/2}$ is the Lorentz factor of the shock.
The cosmic microwave background (CMB), which arranges
$u_{\rm ph}\simeq 4\times 10^{-13}\Gamma^{2}~{\rm ergs~cm^{-3}}$,
plays a minor role in determining $\gamma^{\ast}$.
These EC scatterers would uniformly pervade at the concerned
region, such that $u_{\rm ph}={\rm const}$ is assumed, in contrast
to the inhomogeneous $|{\bf B}|$ (which is dependent on $\lambda$).

In Figure~1, for a dominant stellar emission level of
$u_{\rm ph}=10^{-9}~{\rm ergs~cm^{-3}}$ and a promising
parameter set of $\beta=4.3$ (for the reasoning, see \S~2.2),
$\beta^{\prime}=5/3$, and ${\tilde l}D=110~{\rm pc}$,
we show the $\lambda$ dependence of $\gamma^{\ast}$ for
a fixed $b=10^{-3}$, given $B_{m}=100~{\rm\mu G}-10~{\rm mG}$
as a parameter ({\it top}), and for $B_{m}=1~{\rm mG}$,
given $b=10^{-4}-10^{-2}$ ({\it middle}).
Here we have set the compression ratio to $r=3.2$
\citep[for $p\simeq 2.3$, inferred from the radio-optical index
of $\alpha=0.67$ at knot\,A;][]{perlman01}, the shock speed to
$U=0.5c$ \citep[as compared to the pattern speeds at knot\,A;][]
{biretta95}, and the inclination angle of the local mean field
to the maximum: $\cos^{-1}(U/c)=60^{\circ}$
\citep[e.g.,][]{owen89,capetti97,heinz97}.
It is found that in any case, $\gamma^{\ast}$ appears
to take the peak value at a certain value of $\lambda=\lambda_{c}$,
and as seen in Figure~1~({\it top}), both
$\gamma_{c}^{\ast}\equiv\gamma^{\ast}|_{\lambda=\lambda_{c}}$
and $\lambda_{c}$ decrease as $B_{m}$ increases.
For the electrons bound to the small filaments with
a size of $\lambda<\lambda_{c}$, the $\gamma^{\ast}$
value is determined by the simple relation of
$r_{g,e}(\propto\gamma^{\ast}/|{\bf B}|)\sim\lambda$
that yields $\gamma^{\ast}\propto\lambda^{(\beta+1)/2}$,
while for the electrons bound to the large filament with
$\lambda>\lambda_{c}$, a radiative loss limits the acceleration.
For the former spatial limit, the $\gamma^{\ast}$-value is
independent of $b$, as seen in Figure~1~({\it middle}).
As for the latter temporal limit, in general, the balance between
the mean acceleration time, which has an approximate dependence of
$\propto r_{g,e}(\lambda/r_{g,e})^{\beta^{\prime}-1}$
\citep{honda04b}, and either the synchrotron cooling time,
$\propto(\gamma^{\ast}|{\bf B}|^{2})^{-1}$ or the IC time,
which has dependencies of $\propto(\gamma^{\ast})^{-1}$
(in the Thomson limit) and $(\ln\gamma^{\ast})^{-1}$
(in the high-energy limit of the Klein-Nishina regime),
yields the scaling of the form of
$\gamma^{\ast}\propto\lambda^{-\mu}$, where
\begin{equation}
\mu=\left\{
\begin{array}{ll}
\mu_{\rm syn}=\left[\left(\beta+1\right)\beta^{\prime}-2\right]
   /\left[2\left(3-\beta^{\prime}\right)\right],\\
\mu_{\rm T}=\left[\left(\beta+1\right)\left(\beta^{\prime}-1\right)
   -\left(\beta-1\right)\right]
   /\left[2\left(3-\beta^{\prime}\right)\right],\\
\mu_{\rm KN}\simeq\mu_{\rm T}[\left(3-\beta^{\prime}\right)
   /\left(2-\beta^{\prime}\right)],\\
\end{array}
\right.
\label{eq:1}
\end{equation}
\noindent
respectively.
In the displayed cases, the shortest loss time is that
of the synchrotron cooling, such that the scaling with
$\mu=\mu_{\rm syn}=2.6$ appears, although for a
less likely case in which both $B_{m}$ and $b$ are
small, an IC limit with $\mu\sim\mu_{\rm KN}=0.35$
can appear (not shown in the figure).

In Figure~1~({\it bottom}), for the convincing parameters
(at knot\,A) of $B_{m}=4.59~{\rm mG}$ (equal to the upper limit
suggested by \citealt{owen89}) and $b=10^{-3}$ (see \S~2.2),
we explicitly show the discrimination between the spatial
({\it dotted line}) and temporal ({\it solid curve}) limits,
indicating the DSA tracks for gyrating electrons trapped in
the filament with $\lambda<\lambda_{c}=1.7~{\rm pc}$
({\it solid arrows}) and $\lambda>\lambda_{c}$
({\it dotted arrows}).
It is also demonstrated that in a high-energy regime and
the possible frequency range of the target photon field
(see \S~3.1), the Klein-Nishina effects reduce the IC loss,
fading the Thomson regime with $\mu_{\rm T}=0.087$
({\it thin dot-dashed line}); as a result, the synchrotron
cooling solely contributes to determine $\gamma^{\ast}$ for
the temporal limit regime, yielding a single scaling with
$\mu_{\rm syn}$ ($=2.6$, without $\mu$-flip; {\it solid curve}).
In any case, one can identify a critical Lorentz factor of
$\gamma_{c}^{\ast}=7.6\times 10^{9}$ at which
all minor electrons bound to the small filaments with
$\lambda<\lambda_{c}$ can escape from them (``escape tracks'')
to spaciously meander in a forest of the magnetized filaments.
For the escaping electrons, more efficient energization
is anticipated, as explained in \S~3.1, wherein the
critical energy $\gamma_{c}^{\ast}mc^{2}$ is pertinently
compared to the injection energy for the additional DSA.
We note that the major electrons quasi-secularly
bound to the large filaments with $\lambda>\lambda_{c}$
(``secular-trap tracks'') cannot reach the critical level of
$\gamma_{c}^{\ast}$ because of the synchrotron cooling.

This type of injection associated with the escape from
the gyrotraps (namely, the bound-free transition) has been
referred to as the transition injection \citep{honda05a}.
It should be again noticed that the ``escape'' does {\it not}
mean the leak of energetic electrons from the jet, but
merely the escape from the magnetized fine filaments
with $\lambda<\lambda_{c}$; that is, for the escape
electrons to be additionally accelerated, of course,
they still need to be confined in the shock accelerator.
More details of the additional DSA are discussed in \S~3.1.

\subsection{\it X-Ray Spectral Index Linked to
Turbulent Spectral Indices}

In the low-energy regime below $\gamma_{c}^{\ast}$, the
gyrating electrons bound to the filaments continuously emit
polarized synchrotron photons in an ordinary fashion.
Thus, for the conventional Fermi-based DSA that engenders
the power-law electron distribution (denoted in \S~2.1),
the resulting flux density of synchrotron radiation
conforms to the generic scaling
$F_{\nu}\propto\kappa\nu^{-(p-1)/2}|{\bf B}|^{(p+1)/2}$
\citep{longair94}.
For simplicity, we set $\kappa={\rm const}$ at the moment.
If we recall the scaling of $|{\bf B}|\propto\lambda^{(\beta-1)/2}$,
then it follows that the most dominant contribution to
$F_{\nu}$ comes from the electrons bound to the filaments
with an outer scale of $\lambda\sim{\tilde l}D$.
On the other hand, the familiar scaling of the
synchrotron cutoff above which $F_{\nu}$ exponentially
decays, $\nu^{\ast}\propto(\gamma^{\ast})^{2}|{\bf B}|$
$\nu^{\ast}\propto\gamma^{\ast 2}|{\bf B}|$
(valid for $\gamma^{\ast}\gg 1$), is found to lead to
$\nu^{\ast}\propto\lambda^{-[2\mu-(\beta-1)/2]}$ in the
secular-trap tracks of $\lambda_{c}<\lambda\leq{\tilde l}D$
(cf. Fig.~1, {\it bottom}).
Hence, for values of $\mu>(\beta-1)/4$, which are typical for
$\mu=\mu_{\rm syn}(\beta^{\prime}\neq 1)$, the $F_{\nu}$ spectrum
includes the most dominant component having a lowermost $\nu^{\ast}$,
which is considered to dominantly contribute to establish
the so-called ``break frequency'' \citep[commonly identified
in the optical to soft X-ray band;][]{perlman01,marshall02}.
The property that the less dominant component involves a
higher $\nu^{\ast}$ cooperates to steepen the $F_{\nu}$ spectrum,
in accordance with the X-ray measurement that indicates that the
$F_{\nu}$ level falls below that inferred from extrapolation
of the featureless radio-optical continuum to the X-ray region.
The $F_{\nu}$ drop in the frequency change from
$\nu_{b}\equiv\nu^{\ast}|_{\lambda={\tilde l}D}$ to
$\nu_{c}^{\ast}\equiv\nu^{\ast}|_{\lambda=\lambda_{c}}$
can be estimated as
\begin{equation}
\phi=\left[\lambda_{c}/({\tilde l}D)\right]^{(p+1)(\beta-1)/4}.
\label{eq:2}
\end{equation}
Note that equation~(\ref{eq:2}) is, through
$\lambda_{c}$, dependent on $B_{m}$ and $b$
(see Fig.~1, {\it top} and {\it middle}).\footnote{
Eq.~(\ref{eq:2}) reflects the ratio of the averaged
magnetic intensity of small to large filaments; that is,
$\left\{\left[\int_{\lambda_{\rm min}}^{\lambda_{\rm c}}
\lambda^{\beta}(d\lambda/\lambda^{2})\right]^{1/2}
\left[\int_{\lambda_{c}}^{{\tilde l}D}
\lambda^{\beta}(d\lambda/\lambda^{2})\right]^{-1/2}
\right\}^{(p+1)/2}\simeq\phi$ in a good approximation for
$\lambda_{\rm min}\ll\lambda_{c}\ll{\tilde l}D$, where
$\lambda_{\rm min}$ is the minimum scale of $\lambda$,
which might be compared to the skin depth (\S~3.3).}

In Figure~2, for $(\beta,\beta^{\prime})=(4.3,5/3)$,
given $b=10^{-4}-10^{-2}$, we plot $\nu_{b}$ ({\it top}),
$\nu_{c}^{\ast}$ ({\it middle}), and $\phi$
({\it bottom}) as a function of $B_{m}$ for the
typical value of ${\tilde l}D=110~{\rm pc}$
({\it thick curves}) and a shorter value of
${\tilde l}D=11~{\rm pc}$ ({\it thin curves}).
The shock conditions have been chosen to be the same as
those for Figure~1.
It appears in Figure~2~({\it top}) that $\nu_{b}$ decreases
as $B_{m}$ increases, according to the dependence of
$\nu_{b}\propto B_{m}^{-[4\mu_{\rm syn}/(\beta-1)-1]}=B_{m}^{-2.1}$
in the regime in which the synchrotron cooling definitely
limits the value of $\gamma$.
In the major ranges of $B_{m}$ and $b$, as seen in the
middle panel, $\nu_{c}^{\ast}$ remains almost constant,
since the change of $|{\bf B}|_{\lambda=\lambda_{c}}$
is canceled out by that of $(\gamma_{c}^{\ast})^{2}$
in the synchrotron-limiting regime.
Also, in the corresponding regime, along the dependence of
$\phi\propto\lambda_{c}^{2.8}$, the decrease of $\lambda_{c}$
with increasing $B_{m}$ (Fig.~1, {\it top}) leads to the
decrease of $\phi$, as depicted in the bottom panel of Figure~2.
Note that for a small value of $B_{m}$ (leading to a small
$|{\bf B}|_{\lambda=\lambda_{c}}$), the appearance of the
synchrotron limit can be incomplete for a small $b$ that lowers
$\gamma_{c}^{\ast}$ (Fig.~1, {\it middle}) to weaken the
Klein-Nishina effect; this property yields the peculiar
$B_{m}$ dependencies of $\nu_{c}^{\ast}$ and $\phi$
in the marginal ($B_{m}$, $b$)-domain, as shown
in the middle and bottom panels of Figure~2.
The values of $\nu_{b}$, $\nu_{c}^{\ast}$, and $\phi$ are
likely enhanced for large values of $b$, and, in addition,
$\nu_{b}$ and $\phi$ are further boosted for the small
value of ${\tilde l}D$, although $\nu_{c}^{\ast}$ is
unchanged, except for the IC-limiting regime.
With regard to a comparison with the observational data,
for a confirmed radio-optical $F_{\nu}$ spectrum with index
$\alpha=0.67$ \citep{biretta91,perlman01}, for example,
a value of $\nu_{b}\simeq 7\times 10^{14}~{\rm Hz}$
is expected for a common X-ray spectrum (of the core
and the brightest knots, including knot\,A) consistent
with a simple power law with $\alpha=1.46\pm 0.05$
(\citealt{marshall02}; for alternative, see below).
For the case of ${\tilde l}D=110~{\rm pc}$ \citep{biretta91} and
$B_{m}=4.59~{\rm mG}$ \citep{owen89}, we see in Figure~2~({\it top})
that the $\nu_{b}$-value calls for a fluctuation level of
$b\simeq 1\times 10^{-3}$, consistent with the assumption of
the small-amplitude disturbance (\S~2.1); that is why this
$b$-level has been taken into account for Figure~1~({\it bottom}).

Let us now evaluate the $F_{\nu}$ spectral index,
which is well defined in the transitive frequency
range between $\nu_{b}$ and $\nu_{c}^{\ast}$.
Recalling the canonical index $\alpha_{1}[=(p-1)/2]$ for
$\nu<\nu_{b}$, for heuristic purposes, we give the definition of
$\alpha_{2}\equiv-\Delta\log{F_{\nu}}/\Delta\log{\nu}$, where
$\Delta\log{F_{\nu}}=-\alpha_{1}\Delta\log{\nu}+\log\phi$
and $\Delta\log{\nu}[=\log(\nu_{c}^{\ast}/\nu_{b})]
=[2\mu-(\beta-1)/2]\log({\tilde l}D/\lambda_{c})$.
That is, the index is found to be expressed in the form
\begin{equation}
\alpha_{2}=\frac{p-1}{2}+\frac{p+1}{2}
\left[\frac{4\mu(\beta,\beta^{\prime})}{\beta-1}-1\right]^{-1}
\label{eq:3}
\end{equation}
\noindent
for $\mu>(\beta-1)/4$.
Note that equation~(\ref{eq:3}) is generic, as it is independent
of $\lambda_{c}/{\tilde l}D$ and thereby independent of
any magnetic field parameters (such as $B_{m}$ and $b$).
The second term of the right-hand side that just represents the
steepening of the spectrum is coupled, through $p$, with the
first term, to deviate from $1/2$ ($\alpha_{2}=\alpha_{1}+1/2$)
for a conventional synchrotron model \citep{kardashev62,longair94}.
As seems more reasonable, the $\alpha_{2}$ value can be
compared to the spectral index measured in a X-ray band.

In Figure~3, for $\beta^{\prime}=5/3$, we plot, for convenience,
$\alpha_{2}(\mu=\mu_{\rm syn})$ as a function of $\beta$,
for a given $\alpha_{1}$ in the range of $0.5-0.8$.
Clearly, one can find that for $\alpha_{1}={\rm const}$,
the value of $\alpha_{2}$ monotonically increases
as the value of $\beta$ increases.
As denoted in Appendix~A, by inverting equation~(\ref{eq:3})
we can express $\beta$ as a function of $\alpha_{1}$,
$\alpha_{2}$, and $\beta^{\prime}$.
When assigning the observational indices of $\alpha=0.67$
(in the radio-optical band) and $1.46$ (in the {\it Chandra}
regime of $0.5-7.5~{\rm keV}$) to $\alpha_{1}$ and
$\alpha_{2}$, respectively, equation~(\ref{eq:a1})
with equation~(\ref{eq:a2}) gives $\beta=4.3$ for a
reliable case of $\beta^{\prime}=5/3$ (see also Fig.~3).
This $\beta$-value, somewhat larger than the value of
$\sim 2$ expected in an early stage of the current
filamentation instability \citep{montgomery79}, is
amenable to the updated result of a three-dimensional
kinetic simulation that revealed the marked steepening
of spectrum in a fully developed nonlinear phase
(\citealt{silva03}; see their Fig.~3), reflecting accumulation
of self-organized larger filaments via the violent
coalescence of smaller filaments \citep{honda00a,honda00b}.
In fact, the observational results indicate that the X-ray
spectrum tends to be steepened when departing from the
active galactic core \citep{marshall02}, while the
radio-optical index likely remains constant \citep{perlman01};
for the spectral data sets of, e.g., knot\,{\it HST}-1
$[(\alpha_{1},\alpha_{2})=(0.71,1.29)]$, knot\,D
$(0.69,1.33)$, and knot\,A $(0.67,1.57)$, equation~(\ref{eq:a1})
yields $\beta=2.4$, $2.8$, and $6.6$, respectively.
The gradual increase of $\beta$ (departing from $\sim 2$ at the
innermost knot) might, more or less, reflect the actual
dynamic evolution of a filament cluster involving the coalescence,
which has originally been inspected by \citet{owen89}.

The minor component of electrons bound to the small filaments
with weak $|{\bf B}|$ makes a negligible contribution to
$F_{\nu}$ in the region of $\nu<\nu_{c}^{\ast}$, but
instead, the electrons can escape from the small
filaments, to be additionally accelerated
above $\gamma_{c}^{\ast}$ (\S\S~2.1 and 3.1).
The energetic synchrotron emission from the free
(not persistently bound) electrons tends to be diffusive,
and the spectral index, $\alpha_{3}$, could be close to
$\beta$ \citep[independent of $p$;][]{toptygin87,fleishman06}.
In the case in which around the critical frequency of
$\nu_{c}^{\ast}$ the flux density level of the diffuse
component is not negligibly smaller than that for the
coherent one, e.g. for $\alpha_{3}>\alpha_{2}$ the
synchrotron spectrum is further steepened above
$\nu_{c}^{\ast}$ to extend the gamma-ray tail.
More on this topic is discussed in \S~3.3
and in Appendices~A and B.

\section{ENERGY REBOOST TO THE PeV RANGE\\* AND GAMMA-RAY EMISSION}

Below we consider the additional DSA of the injected electrons
undergoing the {\it nongyrational} motion in the filamentary plasma.
We evaluate the achievable highest energy (\S~3.1)
and the upper limit of the photon frequency (\S~3.2), and
argue a condition for which the high-frequency tail of the
gamma-ray synchrotron spectrum could be extended (\S~3.3).

\subsection{\it Cutoff Lorentz Factor of Energetic-Tail Electrons}

The sufficient condition for which the free electrons can be
additionally accelerated in situ is that the acceleration time
must be shorter than that for the conventional DSA of gyrating
electrons (for the necessary condition, see \S\S~2.1 and 5).
In the high-energy regime compared to high-$\beta$ plasma state,
the free electrons spaciously meander, suffering successive
small deflections by the quasi-static magnetic fields fluctuating
in the zeroth order (sustained by bulk current filaments).
The fields of many fine filaments can be characterized
by the rms strength of $B=(\left<|{\bf B}|^{2}\right>)^{1/2}$,
where $\left<\cdots\right>$ indicates an average on the
transverse two-dimensional plane of the filaments; an
average vector of the local fields likely vanishes in the
interior of the jet such that $|\left<{\bf B}\right>|\ll B$,
whereby the gyroradius can be no longer well defined.
The key kinetic property in this energetic regime is
that the free electrons quickly lose coherence
of the trajectory in the stochastic medium.
For the off-resonant scattering diffusion \citep{honda05a},
the correct ordering of the mean free path is
$\ell_{e}\sim\epsilon^{2}/(e^{2}B^{2}{\tilde l}D)$;
here note that its anisotropy is small, according to
the three-dimensional rms deflection.
The ratio of the mfp to $r_{g,e}$ can be expressed as
\begin{equation}
\frac{\ell_{e}}{r_{g,e}}\sim 10^{-3}
\frac{\gamma}{10^{9}}\frac{10~{\rm \mu G}}{B}
\frac{100~{\rm pc}}{{\tilde l}D}.
\label{eq:4}
\end{equation}
It is evident that there is a wide parameter range in which
$\ell_{e}<r_{g,e}$.\footnote
{Relating to a major difference from the ion kinetics
that was examined in the preceding work \citep{honda04a},
one can find the inequality relation of
$\ell_{e}/r_{g,e}\ll\ell_{i}/r_{g,i}\leq\ell_{p}/r_{g,p}$,
where the subscripts ``$i$'' and ``$p$'' indicate
arbitrary ions and protons, respectively.
In this sense, the quick loss of the coherence in electron
motion is more pronounced than that in ion motion.}
For example, for the ordering of
$B\sim|{\bf B}|_{\lambda\sim\lambda_{c}}\sim 10~{\rm\mu G}$ and
${\tilde l}D=110~{\rm pc}$ \citep{biretta91}, the relation of
$\ell_{e}<r_{g,e}$ is safely satisfied in the range of $\gamma\leq 10^{11}$.
The shorter mfp (in the shock normal direction) is
essential for more rapid back-and-forth across the
shock, resulting in a shorter DSA timescale.

If we allow ${\tilde l}$ to take a value less than unity,
the modified version of the mean acceleration time for
the electrons is found to give the scaling of
\begin{equation}
t_{\rm acc}\simeq 10
\left(\frac{\gamma}{10^{11}}\right)^{2}
\left(\frac{10~{\rm \mu G}}{B}\right)^{2}
\frac{100~{\rm pc}}{{\tilde l}D}
\left(\frac{0.5c}{U}\right)^{2}~{\rm yr}.
\label{eq:5}
\end{equation}
\noindent
The numerical factor of equation~(\ref{eq:5}) is correct
for $\beta=4.3$ and $r=3.2$ (expected at knot\,A), although
it is merely weakly dependent on these parameters.
For, e.g., $B\sim 10~{\rm\mu G}$, ${\tilde l}D=110~{\rm pc}$, and $U=0.5c$,
the acceleration timescale (eq.~[\ref{eq:5}]) is, at
$\gamma=\gamma_{c}^{\ast}(\sim 10^{9}-10^{10})$,
favorably shorter than the critical DSA timescale of
$[r/(r-1)](cr_{g,e}/U^{2})\sim 1-10~{\rm yr}$ for
the Bohm diffusion limit.\footnote{
Concerning the limitation of $t_{\rm acc}$ that depends on
the diffusion coefficient, there is a general constraint
for which the diffusion speed of electrons must be
larger than $U$ \citep{jokipii87,honda05b}.
For nonrelativistic shocks, indeed, this constraint
turns out to be satisfied, as far as the electron mfp
in the shock's normal direction does not greatly exceed
the characteristic length of the density gradient.
Also, equation~(\ref{eq:5}) cannot take unlimitedly smaller values
with decreasing $\gamma$, because the gyrotrapping effects
of cold electrons in the local magnetic fields make the
approximation of $|\left<{\bf B}\right>|\ll B$ worse.
The lower limit of $\gamma$ reflects the energy
required for escape from the local magnetic traps.}
Importantly, this ensures that the sufficient condition
for further energization of the escape electrons with
$\gamma=\gamma_{c}^{\ast}$ (and less) is satisfied.
In particular, the critical energy $\gamma_{c}^{\ast}mc^{2}$ may be
associated with the injection energy for the additional DSA (\S~2.1),
and the conventional DSA of the gyrating electrons in the escape
tracks (illustrated in Fig.~1, {\it bottom}) can be regarded as
the ``preaccelerator'' in the envisagement of the two-step DSA.
Since the shorter $t_{\rm acc}$ of equation~(\ref{eq:5})
can be retained up to the very energetic regime of
$\gamma\sim 10^{11}$ (in conformity with $\ell_{e}<r_{g,e}$),
we expect that two electron populations mediated by the
injection threshold, namely, the major low-energy and
minor high-energy components, will be established.
Recall here that for the Fermi (type I) fashion, the resulting
index $p$ (for the power-law $dn/d\epsilon$ distribution) is
independent of the diffusion coefficient, depending solely on
the scattering center compression ratio $r$ (\S~2.1).
Accordingly, for the present model supposing ubiquitous Fermi
operation in the two energy levels, it is, at least for a
high Mach number case, expected that the power-law distribution
likely has a common $p$ \citep{vainio99,schlickeiser02}.

As usual, the additional DSA of the injected electrons operates
as long as they are confined in the accelerator, where the
electron kinetic energy can increase until $t_{\rm acc}$
asymptotically approaches the radiative loss times.
This spatiotemporal limit truncates the power-law tail of
such a $dn/d\epsilon$ distribution.
We note that the spatial limit may be denoted as
$\ell_{e}\sim D$, instead of $r_{g,e}\sim\lambda$ for the
pre-DSA, and as for the synchrotron loss, the diffuse effects
could be nonnegligible \citep{toptygin87,fleishman06}.
We here take, for convenience, the harmonic average
of the synchrotron and IC cooling times to get
$t_{\rm acc}=t_{\rm syn}[1+a{\tilde\tau}
(\sigma_{\rm KN}/\sigma_{\rm T})]^{-1}$,
where $t_{\rm syn}$ is the diffuse synchrotron time
derived in Appendix~B, ${\tilde\tau}$ is a dimensionless
factor (see eq.~[\ref{eq:b4}]), $a\equiv 8\pi u_{\rm ph}/B^{2}$,
and $\sigma_{\rm KN}/\sigma_{\rm T}$ is the ratio of the
Klein-Nishina and Thomson cross sections
\citep[e.g.,][]{longair92,longair94}.
Note that this time-balance equation does not explicitly
contain $B$, because both $t_{\rm acc}$ and $t_{\rm syn}$
have a common dependence of being proportional to $B^{-2}$.
Solving the spatiotemporal balance equation for $\gamma$,
given ${\tilde l}D$, $U$, $a$, and the characteristic frequency
of target photons in the comoving frame, $\nu_{t}$ (involved in
$\sigma_{\rm KN}$), yields the achievable maximum (cutoff)
Lorentz factor of an accelerated electron, defined as
$\gamma_{c}\equiv\min\left(\gamma|_{t_{\rm acc}=t_{\rm loss}},
\gamma|_{\ell_{e}=D}\right)$.

In Figure~4, for the standard parameter set of
${\tilde l}D=110~{\rm pc}$ and $U=0.5c$ (\S~2), the numerical
solutions, $\gamma_{c}$, are plotted as a function of $\nu_{t}$
in the parameter range of $10^{1}\leq a\leq 10^{6}$.
In any case, we find that the spatial limit plays a
nondominant role in determining $\gamma_{c}$ (thereby,
$\gamma_{c}=\gamma|_{t_{\rm acc}=t_{\rm loss}}$), in contrast
to the previous case of determining $\gamma^{\ast}$ (\S~2.1).
For the larger $a{\tilde\tau}$ value, we see that in the
Thomson limit of $h\nu_{t}\ll mc^2/\gamma_{c}$, where $h$
is the Planck constant, the values of $\gamma_{c}$ take
constants because $\sigma_{\rm KN}/\sigma_{\rm T}\simeq 1$,
while for $h\nu_{t}\gtrsim mc^2/\gamma_{c}$, in which
$\sigma_{\rm KN}/\sigma_{\rm T}<1$, the Klein-Nishina
effect enhances the value of $\gamma_{c}$.
However, $\gamma_{c}$ does not unlimitedly increase, but
eventually saturates, because of the dominance of the
diffuse synchrotron cooling loss, to take a constant value.
For variations of $\nu_{t}$, the allowed domain of the
other variables can be thus written as follows:
\begin{equation}
1\geq\frac{\gamma_{c}}{5\times 10^{10}}
\left(\frac{100~{\rm pc}}{{\tilde l}D}\right)^{1/3}
\left(\frac{0.5c}{U}\right)^{2/3}\geq\frac{1}{(1+a{\tilde\tau})^{1/3}}.
\label{eq:6}
\end{equation}
\noindent
Note that the upper bound and $a{\tilde\tau}\ll 1$
reflect the diffuse synchrotron limit and the lower
bound for $a{\tilde\tau}>1$ reflects the Thomson limit.

In the region in which $B<1~{\rm mG}$, the EC scatterers,
including the starlight and dust emission (\S~2.1),
can lead to $a>1$ for $1\lesssim\Gamma <10$.
As seen in Figure~4, however, in their characteristic
frequencies of $\nu_{t}\sim 10^{14}\Gamma~{\rm Hz}$ (stars)
and $3\times 10^{13}\Gamma~{\rm Hz}$ (dust) the values
of $\gamma_{c}$ turn out to be limited by the diffuse
synchrotron cooling rather than the IC cooling, for the displayed
range of $a\leq 10^{6}$ that would cover the conceivable $a$-values.
This is because the Klein-Nishina effects prolong the
IC timescale beyond the diffuse synchrotron time.
Although the CMB provides a lower target frequency
($\nu_{t}\simeq 2\times 10^{11}\Gamma~{\rm Hz}$)
that tends to fade the Klein-Nishina effects, again
it will play a negligible role in determining $\gamma_{c}$,
because $a<1$; the SSC process is also ignorable, as the
source arranges a lower $u_{\rm ph}$ and higher $\nu_{t}$
compared to those of the above EC sources \citep{stawarz03}.
Consequently, for the standard value of $\beta=4.3$ and another
possible value of $\beta=6.6$ (at knot\,A; \S~2.2), we obtain
$\gamma_{c}\simeq 5\times 10^{10}$ (Fig.~4) and $4\times 10^{10}$,
respectively, exceeding $\gamma_{c}^{\ast}$, and then the
diffuse synchrotron timescales are, for example, for
$B\sim 10~{\rm\mu G}$, estimated as $t_{\rm syn}$
($\sim t_{\rm acc}$) $\sim 2~{\rm yr}$
(eq.~[\ref{eq:5}]) and $1~{\rm yr}$, respectively.
Even if the ${\tilde l}$-value deviates from unity
(${\tilde l}\lesssim 1$), the decrease of these
$\gamma_{c}$-values is not so significant on account of
the dependence of $\gamma_{c}\propto{\tilde l}^{1/3}$.

\subsection{\it Inverse Compton Gamma-Ray Cutoff}

The electrons concomitant with the truncated power-law
distribution scatter the target photons with frequency
$\nu_{t}$ boosting them to higher frequency regions.
Of particular interest is the achievable highest frequency
of the scattered photons, since a significantly large value
of $\gamma_{c}$ is accomplished in the present scenario.
This is the subject of truncation of the gamma-ray spectrum.

In Figure~5 we plot the highest (cutoff) frequency of
the IC radiation (in the comoving frame),
$\nu_{{\rm ic},c}$, as a function of $\nu_{t}$
in the parameter range of $10^{1}\leq a\leq 10^{6}$.
The values of ${\tilde l}D$ and $U$ are chosen
to be the same as those in Figure~4.
In the high-energy and high-frequency region of
$h\nu_{t}\gg mc^{2}/(4\gamma_{c})$, we find that
$\nu_{{\rm ic},c}\simeq\gamma_{c}mc^{2}/h$, while
for $h\nu_{t}\ll mc^{2}/(4\gamma_{c})$, which covers the
Thomson limit, $\nu_{{\rm ic},c}\simeq 4\gamma_{c}^{2}\nu_{t}$.
Note that in a Klein-Nishina regime satisfying
$t_{\rm ic}>t_{\rm syn}$, where $\gamma_{c}$ takes a
value independent of $\nu_{t}$ (Fig.~4), the former
expression of $\nu_{{\rm ic},c}$ ($\propto\gamma_{c}$) is
independent of $\nu_{t}$, as is obviously seen in Figure~5.
The scaling of this upper limit is written in the simple form
\begin{equation}
\nu_{{\rm ic},c}\simeq 6\times10^{30}
\left(\frac{{\tilde l}D}{100~{\rm pc}}\right)^{1/3}
\left(\frac{U}{0.5c}\right)^{2/3}~{\rm Hz}.
\label{eq:7}
\end{equation}
\noindent
Apparently, this is just the case involved in the EC
scenarios due to the stellar and dust emissions
with $\nu_{t}\sim 10^{14}\Gamma~{\rm Hz}$ and
$3\times 10^{13}\Gamma~{\rm Hz}$, respectively
(\S~3.1, Fig.~5), to safely give the ordering of
$\nu_{{\rm ic},c}\sim 10^{30}~{\rm Hz}$
for the reasonable parameter range.
Again we point out the weak ${\tilde l}$ dependence and
the $B$ independence of equation~(\ref{eq:7}), and
that the CMB-based EC and SSC mechanisms will not
greatly change the $\nu_{{\rm ic},c}$-value.
As a consequence, we can claim that the $\nu_{{\rm ic},c}$-value
multiplied by an appropriate Doppler factor, which is of order
unity in the present regime, provides a common cutoff of the
IC radiation from the source.
Concerning the observability of the very energetic photons,
the absorption in extragalactic background light with
subsequent pair production \citep{nikishov62,gould66} will
be significant in spite of the proximity. The relevant
issues, including applicability to more distant radio
galaxies, are discussed in \S~5.

In Figure~6, for ${\tilde l}D=110~{\rm pc}$ and $U=0.5c$,
we present the model $\nu F_{\nu}$ spectra of the
possible IC radiation due to the EC target with
$\nu_{t}=10^{14}~{\rm Hz}$ (stars) and, for convenience,
the synchrotron radiation for the corresponding case of
$\beta=4.3$ ({\it solid curve}), and another case with
$\beta=6.6$ ({\it dotted curve}) as a comparison.
The synchrotron spectra have been best fitted to the observational
data of near-infrared to optical \citep[at the knot\,A shock;][]
{perlman01} and X-ray bands \citep[knot\,A;][]{marshall02}.
As for the IC radiation component, we plot the intrinsic
(source) spectrum for the possible cases of $a=10^{-2}$
({\it dot-dashed curve}) and $10^{-1}$ ({\it dashed curve}),
as accommodated by the data recently observed by the
HESS telescope \citep{beilicke05} that are
considered to reflect a quiescent epoch of the
(unresolved) gamma-ray source.
Note that the chosen $a$-values will sufficiently
reflect the uncertain value of
$u_{\rm ph}\gtrsim 10^{-9}~{\rm ergs~cm^{-3}}$ and
rms strength of the fields that trap the major electrons
emitting the synchrotron photons with the frequency
$\sim\nu_{b}$ (at which $\nu F_{\nu}$ takes the peak).
As seen in Figure~6, below the break frequency of IC
gamma rays, the $\nu$ dependence of $\nu F_{\nu}$ coincides
with that for the synchrotron continuum [i.e.,
$\nu F_{\nu}\propto\nu^{(3-p)/2}=\nu^{0.33}$],
and the spectrum takes the local maximum value of
$\nu F_{\nu}\simeq 4\times 10^{10}~{\rm Jy~Hz}$ for $a=10^{-1}$
at the break frequency of $\sim 10^{25}~{\rm Hz}$.
Above the break, $\nu F_{\nu}$ is significantly suppressed due to
the Klein-Nishina effects, followed by the asymptotic dependence
of $\nu F_{\nu}\propto\nu^{-(p-1)}\ln\nu$ ($\sim\nu^{-1.2}$
for a spectral fit), and ends up with the cutoff at
$\nu_{{\rm ic},c}\simeq 6\times 10^{30}~{\rm Hz}$ (eq.~[\ref{eq:7}]).
We note that changing the $\beta$-value to $6.6$
leads to a slight shift of the cutoff to
$\nu_{{\rm ic},c}\simeq 5\times 10^{30}~{\rm Hz}$
(not shown in the figure).
In sub-GeV to GeV regimes, the curve for $a=10^{-1}$
reaches the sensitivity limit of the LAT instrument
(Large Area Telescope) on board the {\it Gamma-Ray
Large Area Space Telescope} ({\it GLAST}).

In regard to the synchrotron spectra, the cases of $\beta=4.3$
and $6.6$ reproduce, via equation~(\ref{eq:3}), for
$(\alpha_{1},\beta^{\prime})=(0.67,5/3)$,
$\nu F_{\nu}\propto\nu^{-0.46}$ and $\nu^{-0.57}$
in the transitive frequency range that covers
the {\it Chandra} band \citep{marshall02},
giving $\nu_{b}\simeq 7\times 10^{14}$
and $1\times 10^{15}~{\rm Hz}$, respectively; and in turn,
for the self-consistently determined $b$-levels (about
$1\times 10^{-3}$ and $2\times 10^{-3}$, respectively,
for the typical ${\tilde l}D$ and $B_{m}$; \S~2.2),
we numerically find $\nu_{c}^{\ast}\simeq 1\times 10^{21}~{\rm Hz}$
(Fig.~2, {\it middle}) and $2\times 10^{21}~{\rm Hz}$,
as well as $\phi\simeq 1\times 10^{-5}$ (Fig.~2, {\it bottom})
and $3\times 10^{-6}$, respectively.
As seen in Figure~6, for example, for the case of
$a=10^{-1}$, the $\nu F_{\nu}$ levels of the synchrotron
component are, above the frequencies of $7\times 10^{19}$
and $3\times 10^{19}~{\rm Hz}$ (for each of the
$\beta$-value cases), overcome by those of the IC
component, and at $\nu=\nu_{c}^{\ast}$, they are
lower than the IC level by a factor of about
$4\times 10^{-2}$ and $1\times 10^{-2}$, respectively.

\subsection{\it On the Gamma-Ray Tail of the Synchrotron Spectrum}

Regarding the energetic synchrotron radiation above $\nu_{c}^{\ast}$,
we are concerned with the following two possible channels
of the related electron kinetics in actual circumstances:
(1) the additionally accelerated electrons mainly retain
the meandering motion without getting trapped in the filaments,
and (2) the electrons (with energy $\epsilon$) are transported
to be trapped in the larger filaments with the size ($\lambda$)
in the range of $\lambda_{c}<2r_{g,e}(\epsilon)<\lambda$ ($\leq D$)
and are cooled down through the gyration with $r_{g,e}$.
For case 1, the diffuse synchrotron spectrum exhibiting
$F_{\nu,d}\propto\nu^{-\beta}$ can appear in a region of $\nu>\nu_{c}^{\ast}$.
At $\nu=\nu_{c}^{\ast}$, the flux density ratio of the diffuse
to the coherent synchrotron component is found to scale as
\begin{equation}
\left(\frac{F_{\nu,d}}{F_{\nu}}\right)_{\nu=\nu_{c}^{\ast}}
\simeq 2\times 10^{-15}
\left(\frac{10^{-3}~{\rm pc}}{\left<\lambda\right>}\right)^{3.3}
\left(\frac{\gamma_{c}/\gamma_{c}^{\ast}}{10}\right)^{7.3}
\left(\frac{10~{\rm \mu G}}{B}\right)^{3.3}
\label{eq:8}
\end{equation}
\noindent
for $\beta=4.3$ (see eq.~[\ref{eq:b5}] for an arbitrary $\beta$).
For the case reflected in Figure~1~({\it bottom}) and
for the diffuse synchrotron limit (Fig.~4), we have
$\gamma_{c}/\gamma_{c}^{\ast}\simeq 7$ and $B$
($\sim |{\bf B}|_{\lambda=\lambda_{c}}$) $\sim 5~{\mu\rm G}$,
whereupon in equation~(\ref{eq:8}), the continuity
condition of $(F_{\nu,d}/F_{\nu})_{\nu=\nu_{c}^{\ast}}=1$
requires a considerably small size
$\left<\lambda\right>\sim 3\times 10^{-8}~{\rm pc}$
($9\times 10^{10}~{\rm cm}$).
Although the $\left<\lambda\right>$-value is much larger
than the order of magnitude of the skin depth
($\sim 10^{6}~{\rm cm}$ for a plasma density of
$\sim 10^{-1}~{\rm cm^{-3}}$; \citealt{owen90,dimatteo03}),
it is still unclear whether a large number of the
filaments that are much smaller than
${\tilde l}D$ can actually exist or not.
For $(F_{\nu,d}/F_{\nu})_{\nu=\nu_{c}^{\ast}}\ll 1$, the
critical frequency $\nu_{c}^{\ast}$ can be regarded as a cutoff
accompanied by the exponential-like decay of $F_{\nu}$,
and such a case is shown in Figure~6, as an example.

For case 2, in which the orbit stochasity of the rms
deflection becomes incomplete, the diffuse synchrotron
spectrum will be modified such that the emission of the
fundamental mode appears to be preferentially pronounced.
For an estimation of the effective frequency $\nu_{c}$, one can
employ the simple formula ($\propto\gamma_{c}^{2}|{\bf B}|$),
intending to make a comparison with the previously suggested
cutoff for coherent synchrotron radiation (\S~1).
Combining equation~(\ref{eq:6}) with the formula yields
\begin{equation}
1\geq\frac{\nu_{c}}{1\times 10^{24}~{\rm Hz}}
\frac{100~{\rm\mu G}}{|{\bf B}|_{\lambda>\lambda_{c}}}
\left(\frac{100~{\rm pc}}{{\tilde l}D}\right)^{2/3}
\left(\frac{0.5c}{U}\right)^{4/3}
\geq\frac{1}{(1+a{\tilde\tau})^{2/3}}.
\label{eq:9}
\end{equation}
\noindent
In equation~(\ref{eq:9}), the upper limit (or
$a{\tilde\tau}\ll 1$), achievable in the knot\,A
environment (\S~3.1), gives a scaling of
$\nu_{c}\propto|{\bf B}|_{\lambda>\lambda_{c}}
({\tilde l}D)^{2/3}U^{4/3}$, and then, for given values
of ${\tilde l}D=110~{\rm pc}$ and $U=0.5c$, the
expected range is $\nu_{c}\geq 10^{23}~{\rm Hz}$ for
$|{\bf B}|_{\lambda>\lambda_{c}}\geq 10~{\rm\mu G}$.
Because of the large value of $\gamma_{c}$, the value
of $\nu_{c}$ is found to be strikingly larger than
the previous values of $\sim 10^{15}~{\rm Hz}$
\citep{keel88,perez-fournon88,meisenheimer96,stiavelli97},
which could be deduced from a simple DSA model assuming the
resonant scattering diffusion of the electrons confined in a
uniform jet (e.g., \citealt{biermann87} for a parallel shock case).

\section{\it Variability of the Broadband Spectrum}

To complement the model predictions, we give a discussion of
the variability of the broadband spectrum, shown in Figure~6.
In the low-frequency bands associated with synchrotron
radiation, it is expected that the light curves
reflect the spatial inhomogeneity of the interior.
The synchrotron emissions from the escaped
electrons with Lorentz factors above
$\gamma_{e}^{\ast}=[eB_{m}{\tilde l}D/(2mc^{2})]
(\lambda/{\tilde l}D)^{(1+\beta)/2}$
(i.e., above the dotted line in Fig.~1, {\it bottom})
will exhibit the high-frequency variation with a
typical light-crossing time of order $\sim\lambda/c$.
In this regime, the highest (cutoff) frequency of the
power spectrum might be compared to the plasma frequency.
On the other hand, the electrons in the quasi-secular trap
tracks (the shaded regions in Fig.~1, {\it bottom}) cannot
freely cross the filaments, in which, instead of such a variation
with a timescale of $\lambda/c$, the low-frequency variation
due to the shock intermittency that might reflect the activity
of the galactic central engine \citep{biretta95} could appear,
conforming to the $1/f-1/f^{2}$ flicker-noise spectrum
\citep[e.g.,][]{kataoka01}.
The bound electrons that emit a synchrotron photon with
frequency $\nu$ have a Lorentz factor of
$\gamma=[4\pi mc\nu/(3eB_{m})]^{1/2}(\lambda/{\tilde l}D)^{(1-\beta)/4}$.
Equating $\gamma(\lambda,\nu)$ to $\gamma_{e}^{\ast}(\lambda)$,
we find the solution $\lambda$ (as a function of $\nu$), which is
for convenience labeled as $\lambda_{b}$ (the high-frequency
variation relies on the spatioscale of $\lambda<\lambda_{b}$).
The quantity $c/\lambda_{b}\equiv f_{b}$ then characterizes
the frequency that discriminates between the aforementioned
high- and low-frequency regions of the power spectrum
for the $\nu$-band light curve.
This key frequency, analogous to the bend-over frequency
of the broken power spectrum for blazar light curves
(Appendix~C), is found to scale as
\begin{equation}
f_{b}\simeq 2\times 10^{-8}
\left(\frac{10^{15}~{\rm Hz}}{\nu}\right)^{0.14}
\left(\frac{B_{\rm m}}{1~{\rm mG}}\right)^{0.43}
\left(\frac{100~{\rm pc}}{{\tilde l}D}\right)^{0.71}~{\rm Hz}
\label{eq:10}
\end{equation}
\noindent
for $\beta=4.3$ (see eq.~[\ref{eq:c1}] for the generic expression).
Note the very weak $\nu$ dependence of $f_{b}$, in accordance
with the feature of the blazar light curves.
For the parameter set of $B_{m}=4.59~{\rm mG}$ and
${\tilde l}D=110~{\rm pc}$, equation~(\ref{eq:10})
gives $f_{b}\simeq 4\times 10^{-8}~{\rm Hz}$ and
$2\times 10^{-8}~{\rm Hz}$ for $\nu=10^{15}~{\rm Hz}$
(optical/UV) and $10^{18}~{\rm Hz}$ (X-ray), respectively.
We note that for $\beta=6.6$, these $f_{b}$-values
shift down to $5\times 10^{-9}~{\rm Hz}$ and
$3\times 10^{-9}~{\rm Hz}$, respectively.
As a consequence, the model predicts the characteristic
variation with a $1~{\rm yr}$ and $10~{\rm yr}$ timescale
(for $\beta=4.3$ and $6.6$, respectively), longer than a day
timescale for typical blazar jets (Appendix~C; eq.~[\ref{eq:c1}]).

Concerning the MeV/GeV and sub-TeV/TeV gamma-ray light curves,
the high-frequency variation (reflecting the filament size)
cannot appear, as the bound electrons in the quasi-secular
trap tracks dominantly contribute to Comptonize the target photons.
The variation could solely reflect the intermittence
of the shock accelerator in the flicker-noise regime.
In the EC model supposing a stationary $u_{\rm ph}$ value,
it then appears that the gamma-ray variation must correlate
with the variation of the synchrotron spectrum in the
low-frequency (flicker noise) regime below $f_{b}$.
This feature will provide a constraint to explain the data
of the High Energy Gamma Ray Astronomy ({\it HEGRA})
experiments \citep{aharonian03} that seem to reflect
the flaring phase of the source (see Fig.~6).
Even though the other scenarios (\citealt{bai01} for the SSC;
\citealt{protheroe03} and \citealt{reimer04} for the
synchrotron-proton blazar [SPB] model; \citealt{stawarz06a}
for the EC involving nuclear activity) could also be candidates
for explaining the higher flux level and variation,
we can claim that the present simple model will guarantee
the level shown in Figure~6, which is compatible with the
lower levels recently confirmed by HESS \citep{beilicke05},
as well as consistent with the updated Whipple 10~m
telescope upper limit \citep{lebohec04}.

\section{DISCUSSION AND CONCLUSIONS}

In the present conception, the lack of the filaments
with $\lambda<\lambda_{c}$ leads to the failure of the
transition injection followed by the additional DSA.
Moreover, even if fine filaments exist, the large value of
$\left<\lambda\right>$ lowers the $F_{\nu,d}$ level, as is
deduced from equation~(\ref{eq:8}) (or eq.~[\ref{eq:b5}]).
It is, therefore, important to confirm the presence
of such a small scale in the actual jet.
In this aspect, finding $f_{b}$ and the high-frequency extension
of the power spectrum (for the observed light curve; \S~4)
provide the implication of the presence of the small scale
of $\lambda<\lambda_{b}$, and in addition, the truncation
of the power spectrum just reflects the actual smallest
value of $\lambda$, whose lower limit is expected to be
on the order of $\sim 10^{6}~{\rm cm}$ (\S~3.3).
Another feasible method is availing of the property
according to which emission from the electrons
in the stochastic medium could be unpolarized.
The threshold frequency above which the polarization
decreases is estimated as
$\nu_{\rm th}=(3/4\pi)(|e{\bf A}|/mc^{2})^{2}(|e{\bf B}|/mc)$,
where ${\bf B}=\nabla\times{\bf A}$, to give the
corresponding coherence length that scales as
$\lambda\simeq 2\times 10^{-3}
\left(\nu_{\rm th}/10^{15}~{\rm Hz}\right)^{1/2}
\left(10~{\rm\mu G}/|{\bf B}|\right)^{3/2}{\rm pc}$.
Combined with the expression of $|{\bf B}|$ (as a
function of $\lambda$), the scaling is recast to
\begin{equation}
\lambda\simeq 0.6\left(\frac{\nu_{\rm th}}
{10^{15}~{\rm Hz}}\right)^{0.14}
\left(\frac{1~{\rm mG}}{B_{m}}\right)^{0.43}
\left(\frac{{\tilde l}D}{100~{\rm pc}}\right)^{0.71}~{\rm pc}
\label{eq:11}
\end{equation}
\noindent
for $\beta=4.3$.
In particular, for the knot\,A parameters of ${\tilde l}D=110~{\rm pc}$
and $B_{m}=4.59~{\rm mG}$, equation~(\ref{eq:11}) reduces to
$\lambda\simeq 0.3(\nu_{\rm th}/10^{15}~{\rm Hz})^{0.14}~{\rm pc}$,
simply having a weak dependence on $\nu_{\rm th}$; then,
for example, for the optical range of $\nu_{\rm th}$
\citep{perlman99}, we have
$\lambda\sim (2-3)\times 10^{-1}~{\rm pc}$, smaller than
$\lambda_{c}\simeq 2~{\rm pc}$ (or $7~{\rm pc}$ for $\beta=6.6$).
The direct measurement of $\lambda$ seems to be difficult even by
the modern {\it Hubble Space Telescope} with a spatioresolution
of 0.1\arcsec; viz., $7.8~{\rm pc}$ at the distance to M87
\citep[$\sim 16~{\rm Mpc}$;][]{tonry91}, while the
$\lambda_{c}$-value may be comparable to the spatioresolution.

In conclusion, we have given a full argument in favor of
regarding the M87 jet/knot\,A as a hard X-ray and gamma-ray emitter,
owing to the Pevatron electron accelerator comprising
the two-step DSA based on the filamentary jet model.
In a plausible case in which the Alfv\'enic fluctuations
superimposed on the local mean magnetic fields (sustained
by the bulk current filaments) are in a Kolmogorov turbulent
state, the observed X-ray spectral index of $1.46$ (or
$1.57$) \citep{marshall02} has been reproduced for the spectral
index of filamentary turbulence of $\beta=4.3$ ($6.6$), which
is amenable to the results of the kinetic simulations that
solve a long time evolution of filamentation instability.
For the typical parameter set of a shock speed of $0.5c$
\citep{biretta95}, a field inclination angle of $60^{\circ}$
\citep[e.g.,][]{owen89,capetti97,heinz97}. a compression ratio
of $3.2$ \citep[compatible with the radio-optical index of
$0.67$;][]{perlman01}, a filament correlation scale compared
to the jet diameter of $110~{\rm pc}$ \citep{biretta91},
and the magnetic intensity of an outer scale filament of
$4.59~{\rm mG}$ \citep{owen89}, we have self-consistently
determined the Alfv\'enic fluctuation level and found
that the synchrotron cutoff most likely appears at
$1\times 10^{21}~{\rm Hz}$ (for $\beta=4.3$;
$2\times 10^{21}~{\rm Hz}$ for $\beta=6.6$), where
the flux density is, by a factor of $1\times 10^{-5}$
($6\times 10^{-6}$), lower than that inferred from
the extrapolation of the radio-optical continuum.
For the optical/UV and X-ray light curves,
the bend-over frequencies of their power spectra
have been estimated as $4\times 10^{-8}~{\rm Hz}$
($5\times 10^{-9}~{\rm Hz}$) and $2\times 10^{-8}~{\rm Hz}$
($3\times 10^{-9}~{\rm Hz}$), respectively.
The additional DSA incorporated with the transition injection
turns out to require fine filaments with a transverse
size smaller than $2~{\rm pc}$ ($7~{\rm pc}$) to boost the
Lorentz factor of a free electron to $5\times 10^{10}$
($4\times 10^{10}$), independent of the rms magnetic field
strength of the filaments.
The intrinsic cutoff frequency of the conceivable IC (EC and SSC)
radiation is worked out at $6\times 10^{30}~{\rm Hz}$
($5\times 10^{30}~{\rm Hz}$), which just reflects
the uppermost electron energy of $\sim 20~{\rm PeV}$.
It is expected that the MeV/GeV and sub-TeV/TeV gamma-ray
light curves solely reflect the intermittence of the
shock accelerator in the flicker-noise regime.

The present conception for the generation of very energetic
emission from the large-scale jet will constitute a frontier,
complementary to the other scenarios that predict emission
from the unresolved galactic core (\citealt{protheroe03} and
\citealt{reimer04} for the SPB model as a misaligned BL Lac object).
Concerning potential applicability to more distant sources,
if the similarity of blazar variability to the present one
(eq.~[\ref{eq:10}]; Appendix~C) is confirmed in future
observations, the filamentary jet model could be useful
for inferring the detailed structure of blazar jets.
It then follows that the generic modeling can be used
to reinforce an interpretation of the intrinsic $\nu F_{\nu}$
spectra that have been less well theoretically studied so far.
Even for the nearby galaxy M87 ($z=0.00427$), the cosmic infrared
background (CIB) will opaque the gamma rays with energies
in the range $10-100~{\rm TeV}$, although the uncertainty
of the extragalactic CIB level makes it difficult to
accurately determine the optical depth \citep
{coppi97,protheroe00,aharonian01} and thereby the observed
cutoff (the expected range is indicated in Fig.~6).
By contrast, the PeV photons ought to suffer the absorption
dominantly by the CMB, so that their mfp can be predicted
with very high accuracy, provided that the Lorentz invariance
remains even in the ultrahigh energy region.
In any case, the present model spectrum is readily referable
to the radio-loud sources within $\sim 100~{\rm Mpc}$, for
which the extragalactic background (CMB, CIB) light is almost
transparent for the sub-TeV (TeV) photons dominantly carrying
the IC gamma-ray flux, whereas for the more distant sources
the observed gamma-ray cutoffs could shift to a lower
frequency region, down to the lower limit that seems to
appear around $\sim 10~{\rm GeV}$ \citep{kneiske04}.

Relating to this perspective, the pair-cascade
radiation triggered by the photon-photon annihilation
potentially contributes to raising the extragalactic
gamma-ray background level \citep{sreekumar98,strong04}.
When assuming a stellar emission level, the gamma-ray
luminosity apparently depends on the rms magnetic field
strength; that is, the complexity of the filamentary turbulent
state (including a spectral index deviating from that
of classical hydromagnetic turbulence; eq.~[\ref{eq:a1}]).
But the contribution from all Fanaroff-Riley
(FR) type I radio galaxies containing X-ray jets is
expected not to exceed the one from blazars
\citep{kneiske05} as long as the rms strength is,
on average, not much smaller than the equipartition
value (like the case shown in Fig.~6 for $a\leq 10^{-1}$),
along with the arguments invoking the luminosity
function (see \citealt{stawarz06b} for the details).
Particularly, for the smaller power index of the filamentary
turbulence, the rms strength of the fields that trap major electrons
constituting the synchrotron $\nu F_{\nu}$ peak tends to be
comparable to the uppermost value of $\sim 1~{\rm mG}$ for
the outer scale filament, wherein the smaller contribution
of at most $1~\%$ is merely guaranteed.
These arguments are not affected by the shift of the
intrinsic cutoff in the PeV range, as the dominant
contribution commonly owes to the (sub-)TeV photons
constituting the local maximum of $\nu F_{\nu}$.
To verify the predictions, in addition to the survey of
hidden sources by future high-sensitivity detectors,
comprehensive measurements of the filamentary
turbulence and its spectrum as well would be
of powerful use (e.g., \citealt{carilli96}),
bringing on some interesting information, arguably
linked to synchrotron spectra (\S~2.2; Fig.~3).
Modern laboratory experiments of high-intensity laser-matter
interaction, which can reproduce the similar turbulent state
\citep[e.g.,][]{wei04}, might be also available for
resolving the complexities in detail.\\

\appendix
\section{THE EXPRESSION OF THE SPECTRAL INDEX OF\\*
FILAMENTARY TURBULENCE}

For instruction, here we write down the expression of the
power-law spectral index of magnetized filamentary turbulence,
$\beta$, as a function of the indices $\alpha_{1}$,
$\alpha_{2}$ ($>\alpha_{1}$), and $\beta^{\prime}$ ($\neq 1$).
Again, note that $\alpha_{1}$ (for $\nu<\nu_{b}$) and
$\alpha_{2}$ (for $\nu_{b}<\nu<\nu_{c}^{\ast}$) could be
typically compared to the radio-optical and X-ray indices,
respectively, and $\beta^{\prime}$ indicates the power-law
index of the Alfv\'enic fluctuation disturbing the zeroth-order
magnetic field (permeating through the current filaments).
If we make use of equation~(\ref{eq:1}) for the ordinary
case of $\mu=\mu_{\rm syn}(\beta,\beta^{\prime})$,
the inversion of equation~(\ref{eq:3}) readily gives
\begin{equation}
\beta(\alpha_{1},\alpha_{2},\beta^{\prime})
=\frac{x(\alpha_{1},\alpha_{2})+x_{a}(\beta^{\prime})}
{x(\alpha_{1},\alpha_{2})-3x_{a}(\beta^{\prime})},
\label{eq:a1}
\end{equation}
\noindent
\begin{equation}
x(\alpha_{1},\alpha_{2})=\frac{\alpha_{1}+1}{\alpha_{2}-\alpha_{1}},~~~~~
x_{a}(\beta^{\prime})=\frac{\beta^{\prime}-1}{3-\beta^{\prime}}.
\label{eq:a2}
\end{equation}
\noindent
Particularly, for a likely case of $\beta^{\prime}=5/3$
(reflecting the Kolmogorov turbulence), we have $x_{a}=1/2$.

In conjunction with the synchrotron spectrum, in the
very high frequency range of $\nu>\nu_{c}^{\ast}$,
the $F_{\nu}$ spectrum could have an index of
$\alpha_{3}=\beta$ (\S\S~2.2 and 3.3; eq.~[\ref{eq:b1}]).
As is, it may be also instructive to clarify what condition would
determine the softening ($\alpha_{3}>\alpha_{2}$) or hardening
($\alpha_{3}<\alpha_{2}$) of the spectrum above $\nu_{c}^{\ast}$.
The algebraic manipulations lead to the {\it Ansatz} that the
relation $\beta>\alpha_{2}$ ($\beta<\alpha_{2}$) is obtained
for the condition $\alpha_{2}>\alpha_{2}^{\dagger}$
($\alpha_{2}<\alpha_{2}^{\dagger}$), where the critical
index $\alpha_{2}^{\dagger}$ can be expressed as a
function of $\alpha_{1}$ and $\beta^{\prime}$ in the form
\begin{equation}
\alpha_{2}^{\dagger}(\alpha_{1},\beta^{\prime})
=\frac{x_{b}(\alpha_{1},\beta^{\prime})
+\sqrt{x_{b}^{2}(\alpha_{1},\beta^{\prime})
-12\alpha_{1}x_{a}(\beta^{\prime})x_{c}(\beta^{\prime})}}
{6x_{a}(\beta^{\prime})},
\label{eq:a3}
\end{equation}
\noindent
where the abbreviations are
$x_{b}=\alpha_{1}(1+3x_{a})+x_{c}$ and $x_{c}=1-x_{a}$.
In the special case that satisfies the equality of
$\alpha_{2}=\alpha_{2}^{\dagger}$, we have the
relation of $\alpha_{2}=\beta$ leading to a
common index of $\alpha_{3}=\alpha_{2}$.
For example, given $\alpha_{1}=0.67$ \citep[at knot\,A;][]{perlman01},
equation~(\ref{eq:a3}) for $\beta^{\prime}=5/3$
provides $\alpha_{2}^{\dagger}\simeq 1.3$, whereby the
comparison with the index of $\alpha_{2}=1.46\pm 0.05$
to $1.57\pm 0.10$ at the corresponding point
\citep{marshall02} leads to $\alpha_{3}>\alpha_{2}$.\\

\section{THE COOLING TIMESCALE AND FLUX DENSITY LEVEL OF THE\\*
DIFFUSE SYNCHROTRON RADIATION}

On the theoretical basis by \citet{toptygin87}, we derive the
cooling time of the diffuse synchrotron radiation of additionally
accelerated electrons in the magnetized filamentary plasma.
The emissivity of a single electron by the diffuse
synchrotron radiation can be expressed as
\begin{equation}
j(\omega,\gamma)=\frac{2^{\beta}\left(\beta-1\right)
\left(\beta^{2}+7\beta+8\right)}
{\beta\left(\beta+2\right)^{2}\left(\beta+3\right)}
\frac{e^{2}}{c}\frac{\omega_{\rm st}^{2}}{\omega_{0}}
\left(\frac{\gamma^{2}\omega_{0}}{\omega}\right)^{\beta}
\label{eq:b1}
\end{equation}
\noindent
for $\omega$ ($=2\pi\nu$) $\gg\gamma^{2}\omega_{0}$,
where $\omega_{\rm st}\equiv eB/(mc)$ and
$\omega_{0}\equiv 2\pi c/\left<\lambda\right>$ \citep{fleishman06}.
In general, the temporal decay of the electron kinetic energy
conforms to $-(d\epsilon/dt)=\int d\omega j(\omega,\epsilon)$,
where $\epsilon=\gamma mc^{2}$ \citep{longair94}.
For convenience, here we set the lower limit of the $\omega$ integral
to $\gamma^{2}\omega_{0}$ and introduce the variable
transformation of $\omega/(\gamma^{2}\omega_{0})=\xi$.
Then we have
\begin{eqnarray}
\int_{\gamma^{2}\omega_{0}}^{\infty}
d\omega\frac{\omega_{\rm st}^{2}}{\omega_{0}}
\left(\frac{\gamma^{2}\omega_{0}}{\omega}\right)^{\beta}
&=&\gamma^{2}\omega_{\rm st}^{2}
\int_{1}^{\infty}\xi^{-\beta}d\xi \nonumber \\
&=&\frac{1}{\beta-1}\gamma^{2}\omega_{\rm st}^{2}.
\label{eq:b2}
\end{eqnarray}
\noindent
Making use of equation~(\ref{eq:b2}), we obtain the
temporal evolution equation of $\epsilon$ as follows:
\begin{equation}
-\left(\frac{d\epsilon}{dt}\right)
=\frac{4}{3}{\tilde\tau}^{-1}(\beta)\sigma_{\rm T}cu_{m}
\left(\frac{\epsilon}{mc^{2}}\right)^{2},
\label{eq:b3}
\end{equation}
\noindent
where $\sigma_{\rm T}$ is the Thomson cross section,
$u_{m}\equiv B^{2}/8\pi$, and
\begin{equation}
{\tilde\tau}(\beta)\equiv\left(6\pi\right)^{-2}
\frac{\beta\left(\beta+2\right)^{2}\left(\beta+3\right)}
{2^{\beta}\left(\beta^{2}+7\beta+8\right)}.
\label{eq:b4}
\end{equation}
\noindent
The solution of equation~(\ref{eq:b3}) is found to be
$\epsilon(t)=\epsilon_{0}\left(1+t/t_{\rm syn}\right)^{-1}$,
where $\epsilon_{0}$ ($=\gamma_{0}mc^{2}$) $\equiv\epsilon(t=0)$.
Here $t_{\rm syn}$ defines the characteristic decay time for
the diffuse synchrotron radiation, which can be expressed as
$t_{\rm syn}={\tilde\tau}{\hat t}_{\rm syn}$, where
${\hat t}_{\rm syn}\equiv 3mc/(4\sigma_{\rm T}u_{m}\gamma_{0})$.
Note that the form of ${\hat t}_{\rm syn}$ appears to be
the same as that for the normal synchrotron cooling
time \citep{longair94}, although $u_{m}$ now includes the
rms field strength $B$, instead of a mean field strength.
For values of, e.g., $\beta=4.3$ and $6.6$,
equation~(\ref{eq:b4}) gives ${\tilde\tau}=3.1\times 10^{-3}$
and $1.4\times 10^{-3}$, respectively.

In terms of the diffuse synchrotron radiation spectrum of
$J_{d}(\omega)=\int j(\omega,\epsilon)\kappa\epsilon^{-p}d\epsilon$,
we pay attention to the ratio to the ordinary synchrotron spectrum
[e.g., eq.~[18.49] for $J(\omega)$ in \citealt{longair94}].
Particularly, at $\omega_{c}^{\ast}=2\pi\nu_{c}^{\ast}$
the ratio can be written in the following form:
\begin{eqnarray}
\left[\frac{J_{d}(\omega)}{J(\omega)}\right]_{\omega=\omega_{c}^{\ast}}
&=&\frac{2^{-p/2}}{27\pi g(p)}
\frac{\left(2/3\right)^{\beta-(1/2)}\left(\beta-1\right)}
{\left(2\beta+p+1\right){\tilde\tau}(\beta)}
\left(\frac{\omega_{\rm st}}{\omega_{0}}\right)^{-(\beta-1)}
\left(\frac{\gamma_{c}}{\gamma_{c}^{\ast}}\right)^{2\beta-p+1} \nonumber \\
&=&\left(\frac{F_{\nu,d}}{F_{\nu}}\right)_{\nu=\nu_{c}^{\ast}},
\label{eq:b5}
\end{eqnarray}
\noindent
where the dimensionless function $g(p)$ including
gamma functions is identical with the $a(p)$ of
equation~(18.49) and Table~18.2 in \citet{longair94}.
It is noted that in the derivation of equation~(\ref{eq:b5}),
the relation of $\omega_{\rm st}/\omega_{c}^{\ast}=(2/3)
(\gamma_{c}^{\ast})^{-2}$, which is valid for the ordering
$B\sim|{\bf B}|_{\lambda=\lambda_{c}}$, has been used.
For a value of $\beta$ that is not small, the factor
$(\omega_{\rm st}/\omega_{0})^{-(\beta-1)}$
can be considerably small (because typically
$\omega_{\rm st}/\omega_{0}\gg 1$), such that the flux density
ratio, $(F_{\nu,d}/F_{\nu})_{\nu=\nu_{c}^{\ast}}$, is much
smaller than unity (see eq.~[\ref{eq:8}] for $\beta=4.3$).
By contrast, for a not-so-small value of equation~(\ref{eq:b5})
that might be realized at inner knots (such as knot\,D)
with a smaller value of $\beta\sim 2-3$ (\S~2.2), the
energetic tail of the diffuse synchrotron radiation can appear
above $\nu_{c}^{\ast}$, to exhibit $F_{\nu,d}\propto\nu^{-\beta}$.\\

\section{THE BEND-OVER FREQUENCY OF THE VARIATION
POWER\\* SPECTRUM FOR BLAZAR GEOMETRY}

We check the validity of equation~(\ref{eq:10}) by comparing
with the corresponding frequency of the variation power spectrum
for blazar jet geometry with a narrow viewing angle.
This is associated with (1) application of the filamentary
jet model to blazar jets and (2) verification of the idea
regarding the FR~I radio jet as a misaligned BL Lac object
\citep{tsvetanov98}; that is, unifying astrophysical jets.
If we take the relativistic beaming of the filamentary bulk flow into account,
the synchrotron frequency is boosted by the beaming factor $\delta$
(i.e., $\nu\propto\delta\gamma^{2}|{\bf B}|$ in the laboratory frame),
such that the expression of $\gamma(\lambda,\nu)$ denoted in \S~4
should be also corrected, to be multiplied by $\delta^{-1/2}$.
Note that the expression of $\gamma_{e}^{\ast}(\lambda)$
(derived from the spatiolimit) is unchanged.
For electron acceleration by fully relativistic shocks
(with the Lorentz factor $\Gamma$), we expect that a
strong anisotropy of the momentum distribution function
is established, such that the accelerated electrons are
preferentially directed in the narrow cone angle of order
$\sim 1/\Gamma$ (in the bulk rest frame), regardless of the
complexity of turbulent scatterers \citep{achterberg01}.
This implies that the transverse light-crossing
will be delayed by a factor of $\Gamma$.
Therefore, the bend-over frequency in the laboratory frame
may be redefined as
$f_{b}\equiv\delta[c/(\Gamma\lambda_{b})]$, where
$\lambda_{b}$ denotes the solution $\lambda$ of the equality
$\delta^{-1/2}\gamma(\lambda,\nu)=\gamma_{e}^{\ast}(\lambda)$
(see also \S~4).
As a consequence, we find that $f_{b}$ can be expressed as
\begin{equation}
f_{b}=\frac{c\delta}{{\tilde l}D\Gamma}
\left[\frac{3}{4\pi}\left(\frac{e}{mc}\right)
\frac{\delta B_{m}}{\nu}\right]^{2/(3\beta+1)}
\left(\frac{eB_{m}{\tilde l}D}{2mc^{2}}\right)^{4/(3\beta+1)}.
\label{eq:c1}
\end{equation}
\noindent
Particularly, setting $\delta=\Gamma=1$ and $\beta=4.3$
yields the scaling of equation~(\ref{eq:10}).

For the bulk Lorentz factor of $\lesssim\Gamma$, we have
the relation of $\delta\lesssim\Gamma$ for blazar geometry,
and then the scaling of equation~(\ref{eq:c1}), e.g., for
$\beta=4.3$, appears to be equation~(\ref{eq:10}) merely
multiplied by the factors of $\delta/\Gamma$ ($\lesssim 1$)
and $\delta^{2/(3\beta+1)}=\delta^{0.14}$.
For the typical blazar parameters of $B_{m}=10^{-1}~{\rm G}$,
${\tilde l}D=10^{-2}~{\rm pc}$, and $\delta=10$, we obtain
the value of $f_{b}\lesssim 5\times 10^{-5}~{\rm Hz}$ for
$\nu=10^{18}~{\rm Hz}$ (X-ray light curve), which is
reasonably consistent with the observational data
\citep[e.g., $f_{b}\simeq 3\times 10^{-5}~{\rm Hz}$
for Mrk\,501;][]{kataoka01}.
The observational trend of the $\nu$ independence of
$f_{b}$ can be also inferred from equation~(\ref{eq:c1})
for conceivable $\beta$-values.
According to these, the use of equation~(\ref{eq:10}) will be
justified for the concerned case of a not-so-narrow viewing
angle and a weak relativistic shock \citep{biretta95,lobanov03}.\\

\clearpage
\section*{FIGURE CAPTIONS}

\figcaption{
Lorentz factor of an electron achievable for a
conventional DSA $\gamma^{\ast}$, vs. the transverse size
of a magnetized filament $\lambda$ for a fixed
$b$ and some given values of $B_{m}$ ({\it as labeled};
{\it top}) and
vice versa ({\it middle}), and for $b=10^{-3}$ and
$B_{m}=4.59~{\rm mG}$ \citep[{\it bottom};][]{owen89},
indicating the
regimes where $\gamma^{\ast}$ is determined by
the spatial ({\it thick dotted line}) and temporal limits
(synchrotron loss; {\it thick solid curve}).
Here $b$ is the energy density ratio of the Alfv\'enic
fluctuation/local mean magnetic field, $B_{m}$ is the
magnetic intensity of the filament with $\lambda$
reaching a correlation length compared to the
jet diameter, $D=110~{\rm pc}$ at knot\,A
\citep[{\it thin long-dashed line};][]{biretta91},
and more on the DSA parameters (at the knot\,A region)
is given in the text.
The axes are common in the figures.
Note that the solid curve in the top panel is
the same as that in the middle panel.
In the bottom panel, the thin dot-dashed curve indicates the
(pseudo)limit due to the Thomson scattering, which is removed by the
Klein-Nishina effect, and the accelerator is active in the
shaded domain, wherein the solid arrows represent the pre-DSA
of the electrons bound to the filaments with $\lambda<\lambda_{c}$,
leading to the escape (bound-free transition) entailed in the
injection of additional DSA, and the dotted arrows represent
the energization that is knocked down by the synchrotron loss
before the escape, namely, the tracks of secular traps in the
filaments with $\lambda>\lambda_{c}$.
[{\it See the electronic edition of the Journal for a
color version of this figure.}]}

\figcaption{
Dependencies of the break $\nu_{b}$ ({\it top}),
the critical frequency $\nu_{c}^{\ast}$ ({\it middle}),
and the flux density reduction factor $\phi$
(defined in eq.~[\ref{eq:2}]; {\it bottom}) on $B_{m}$,
for $b=10^{-4}-10^{-2}$ ({\it as labeled}) and the
correlation length of filaments compared to $D$ ($=110~{\rm pc}$;
{\it thick curves}) and $0.1D$ ({\it thin curves}).
Here the other DSA parameters are the same as those for Fig.~1.
The horizontal axis is common in the figures.
Note that in the middle panel, the thick solid and
dot-dashed lines overlap the thin ones, and in the
bottom panel, the thick dot-dashed line overlaps
the thin dotted line.}

\figcaption{
Transitive synchrotron spectral index $\alpha_{2}$
(eq.~[\ref{eq:3}]) vs. the power-law spectral index of
filamentary turbulence $\beta$ for the given canonical
index $\alpha_{1}$ as a parameter ({\it as labeled}).
In connection with observed synchrotron spectra, $\alpha_{1}$
and $\alpha_{2}$ can be typically compared to the radio-optical
and X-ray spectral indices, respectively.
For further explanation, see the text.}

\figcaption{
Cutoff Lorentz factor of an additionally
accelerated electron $\gamma_{c}$ vs. the target photon frequency
$\nu_{t}$, for the energy density ratio of target photons/magnetic
fields in the range $a=10^{1}-10^{6}$ ({\it as labeled}).
Here the parameters relevant to shock and filamentary turbulence
are the same as those for Fig.~1 (see also the text).
The dotted line indicates $\epsilon=\gamma_{c}h\nu_{t}/(mc^{2})=1$:
the region of $\epsilon\ll 1$ reflects the Thomson limit,
in which the values of $\gamma_{c}$ take constants,
while $\epsilon\gtrsim 1$ indicates the Klein-Nishina regime.
An asymptotic value for $a{\tilde\tau}\ll 1$
(${\tilde\tau}\simeq 3\times 10^{-3}$) provides the
upper bound of $\gamma_{c}$ (independent of the
rms magnetic field strength $B$), which is referred
to as the diffuse synchrotron limit.}

\figcaption{
Cutoff frequency of the inverse Compton
gamma-ray radiation $\nu_{{\rm ic},c}$ vs. the target
photon frequency $\nu_{t}$ for $a=10^{1}-10^{6}$
({\it as labeled}), corresponding to the cases shown in Fig.~4.
The ranges of characteristic frequency for the candidate external
Compton (EC) sources of the starlight ("Star"), dust emission ("Dust"),
and the cosmic microwave background ("CMB") are also indicated for the
bulk Lorentz factor of $1-10$ \citep[{\it arrows};][]{stawarz03}.}

\figcaption{
Model $\nu F_{\nu}$ spectrum of synchrotron radiation for
$\beta=4.3$ ({\it solid curve}) and $6.6$
({\it dotted curve}), and for the former, the source spectrum of IC
radiation for the possible cases of $a=10^{-1}$ ({\it dashed curve})
and $10^{-2}$ ({\it dot-dashed curve}) in the knot~A environment.
Here the EC target photon frequency has been supposed to be
$\nu_{t}=10^{14}~{\rm Hz}$, and the other conditions are
the same as those for Figs.~1~({\it bottom}) and 4.
The open squares and diamond indicate the data measured in
the near-infrared to optical \citep{perlman01} and X-ray
\citep{marshall02} bands, respectively, and
the open and filled triangles indicate the gamma-ray
levels detected by HEGRA \citep{aharonian03}
and HESS \citep{beilicke05}, respectively.
In addition, the EGRET \citep{reimer03} and
Whipple 10 m telescope \citep{lebohec04}
upper limits are indicated.
For convenience, the sensitivity limits of a currently operating
Cerenkov telescope \citep[HESS;][]{hofmann01} and future
detectors ({\it GLAST} LAT, \citealt{mcenery04};
{\it New X-Ray Telescope} Hard X-Ray Imager and
Soft Gamma-Ray Detector [{\it NeXT} HXI and SGD],
\citealt{takahashi04}) are also depicted.
The intrinsic cutoff and expected range of the observed
cutoff as a result of the absorption in the CIB
\citep{coppi97,protheroe00,aharonian01} are
indicated by the thick arrow and the thick bar
with thin dotted lines, respectively.
Note that the diffuse synchrotron component that appears in
the high-energy gamma-ray region is of a considerably
lower level. [{\it See the electronic edition of the
Journal for a color version of this figure.}]}

\epsscale{.9}
\clearpage
\plotone{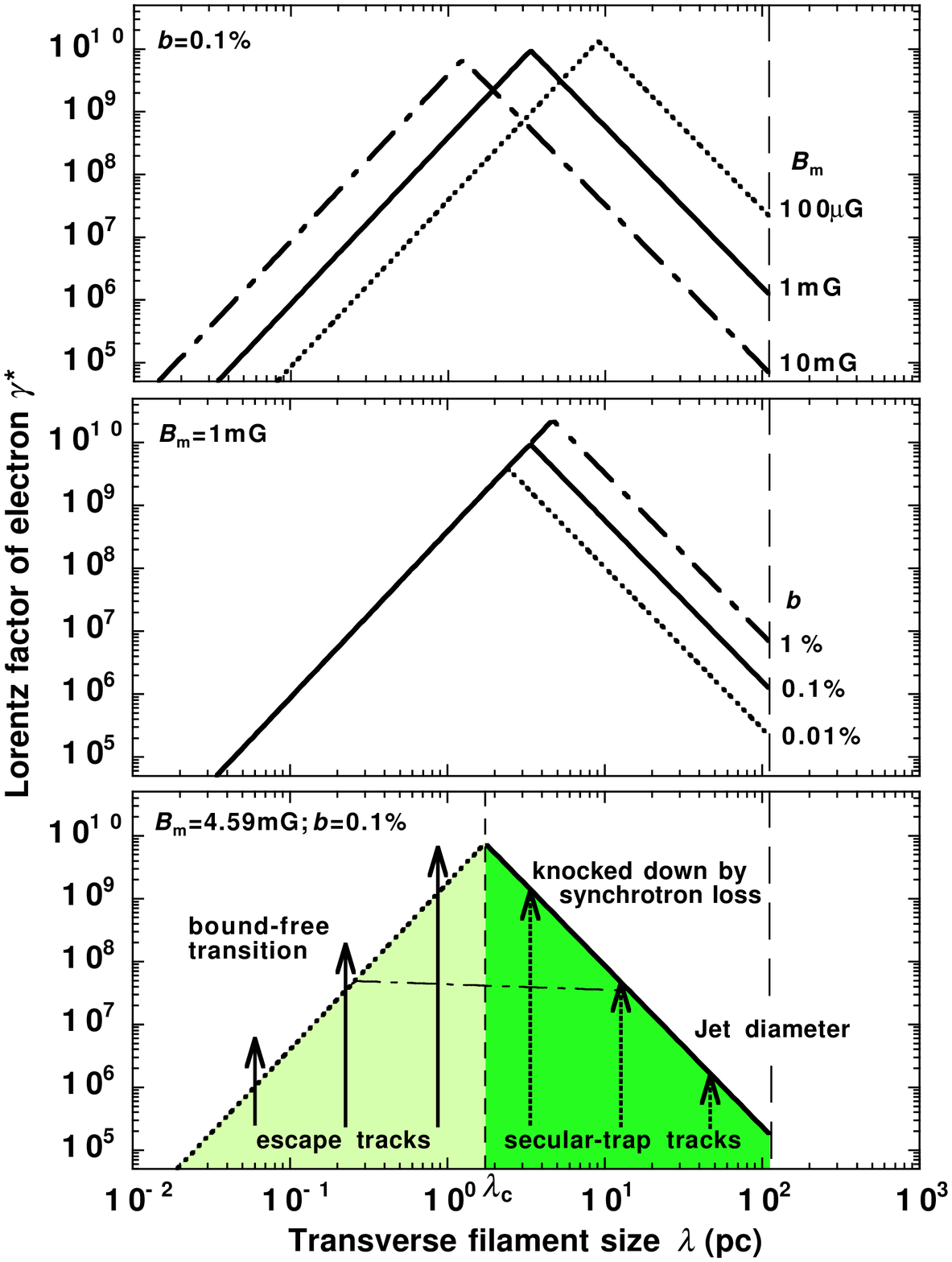}

\clearpage
\plotone{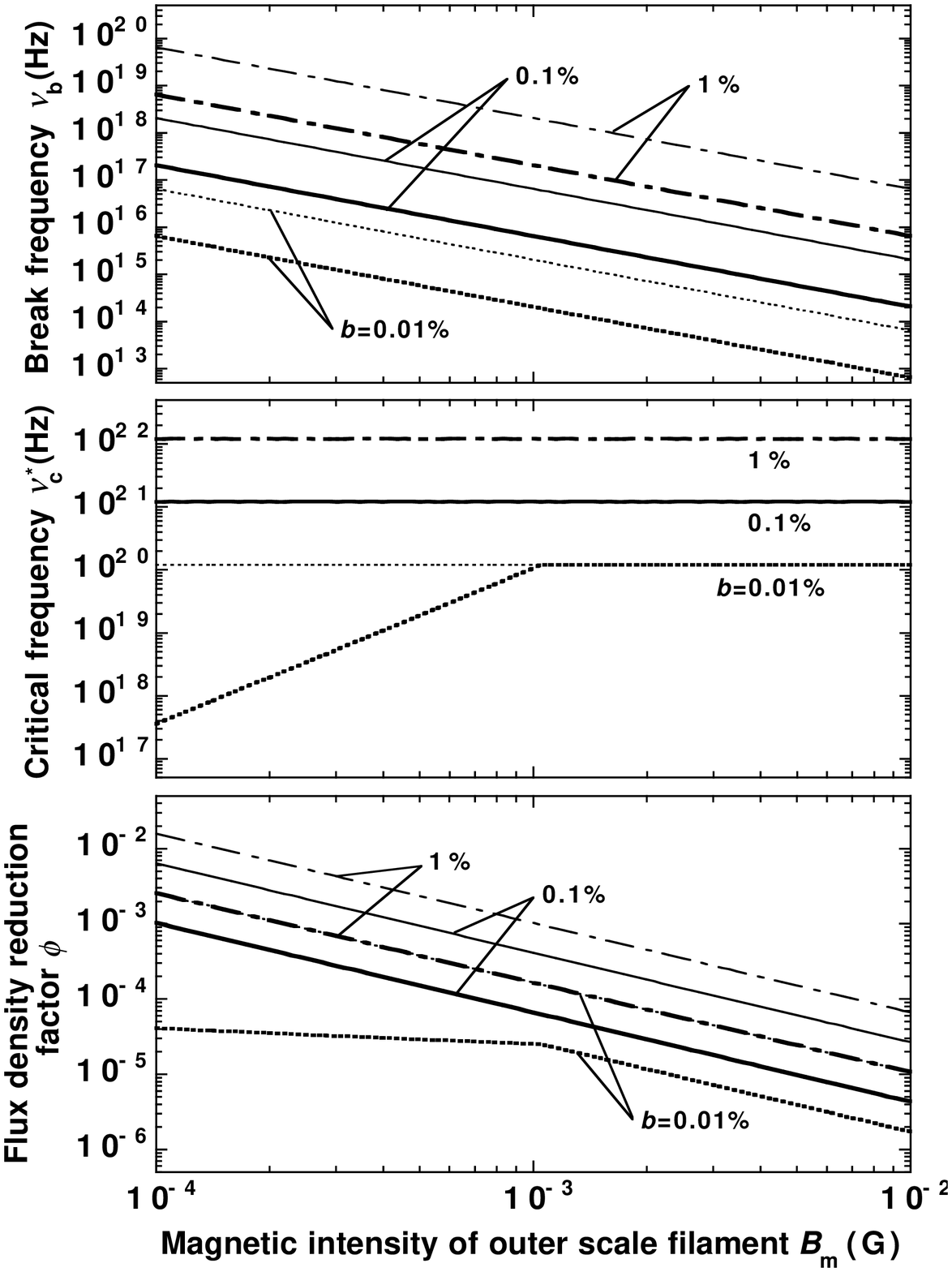}

\epsscale{1}
\clearpage
\plotone{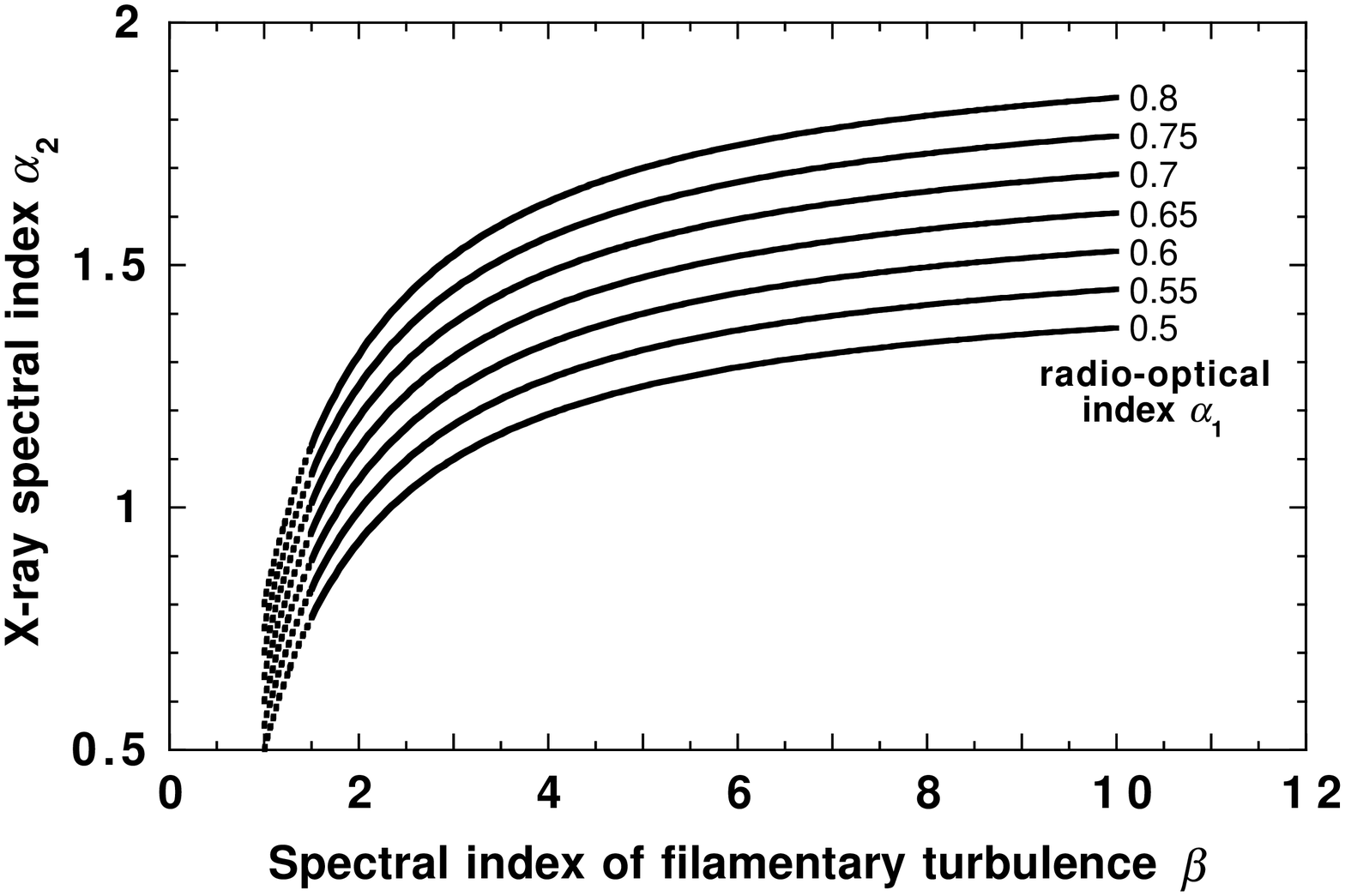}

\clearpage
\plotone{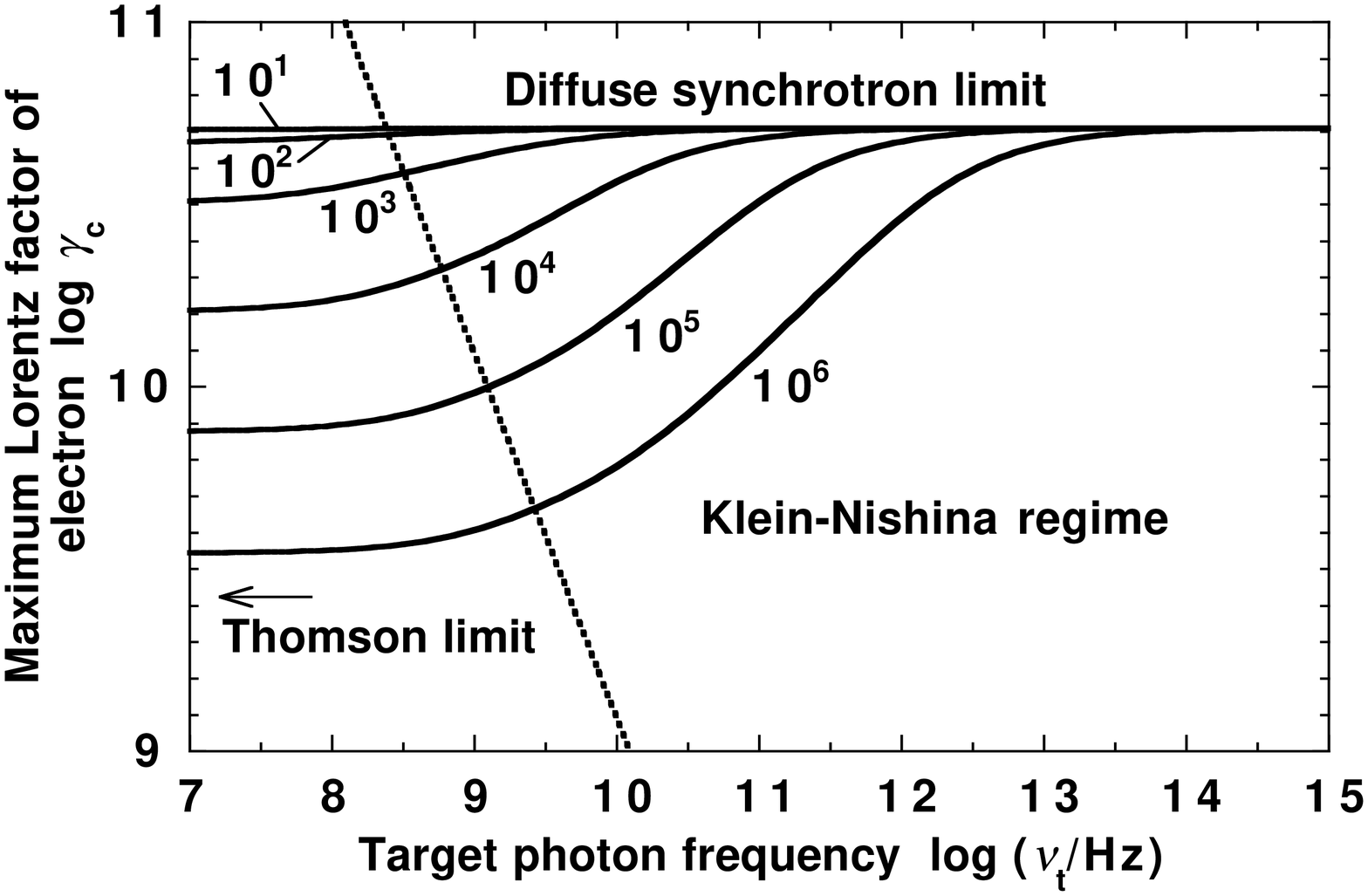}

\clearpage
\plotone{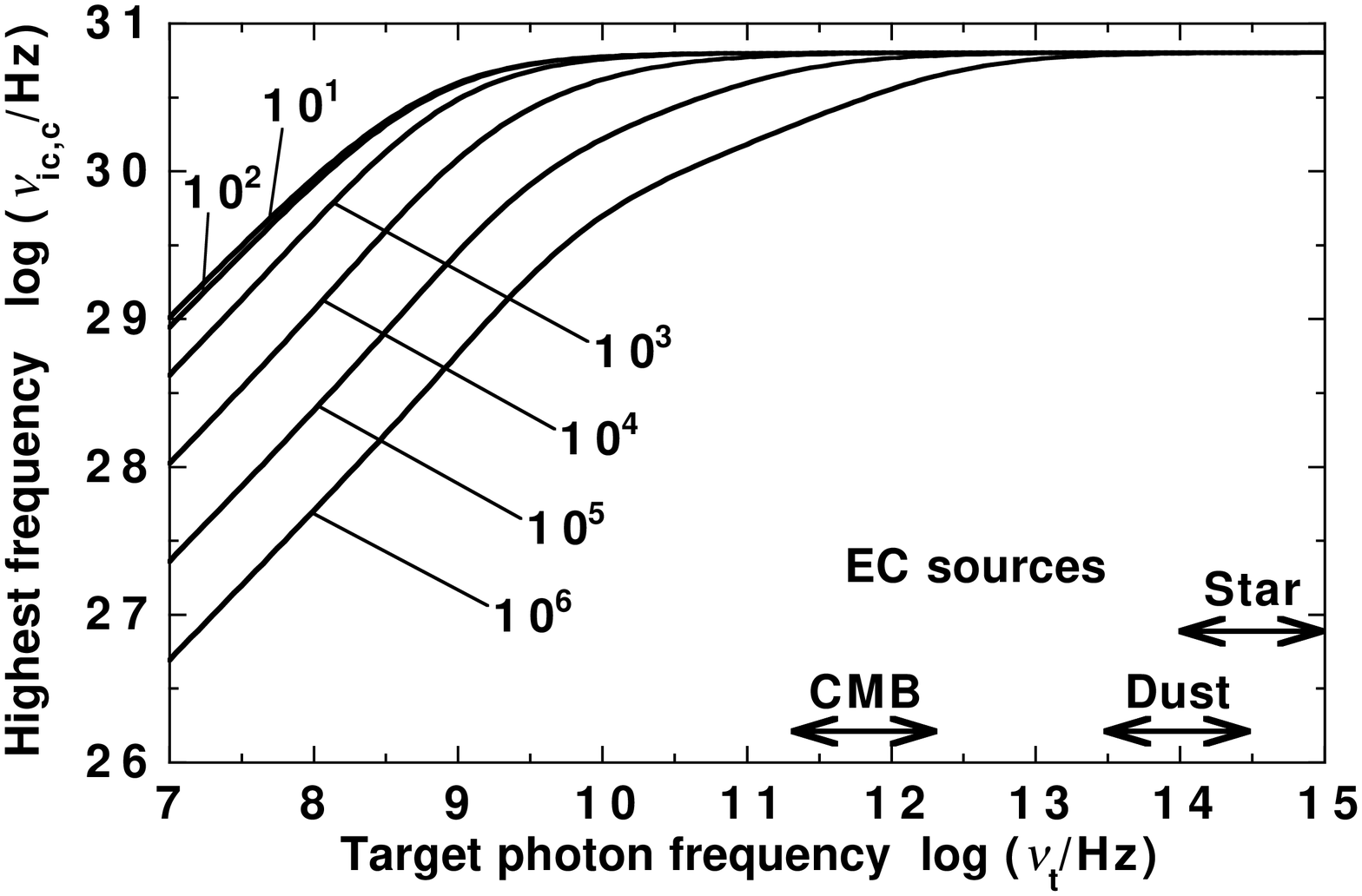}

\clearpage
\plotone{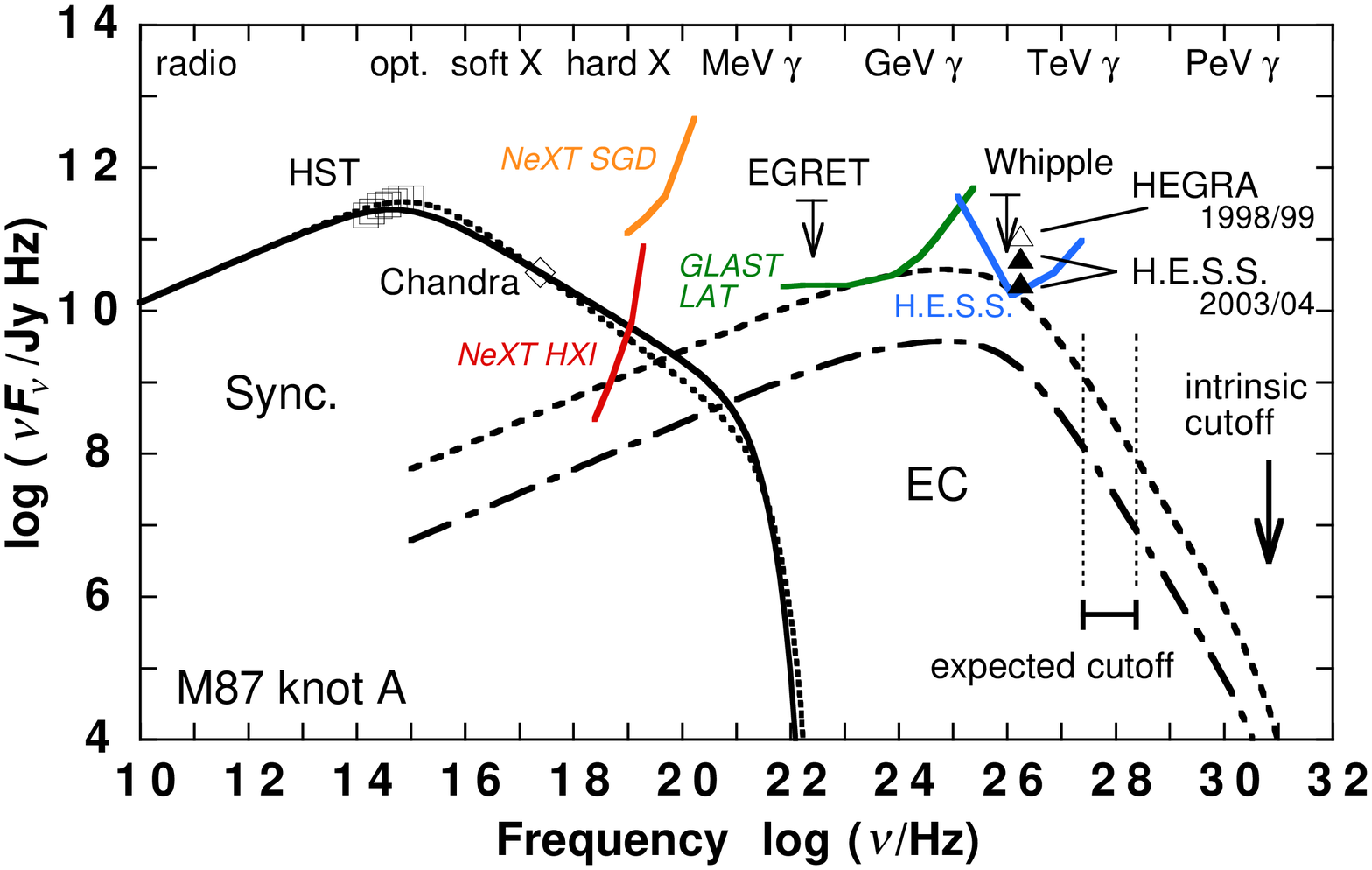}


\begin{thebibliography}{}
\bibitem[Achterberg et al.(2001)]{achterberg01}
Achterberg,~A., Gallant,~Y.~A., Kirk,~J.~G., \& Guthmann,~A.~W.,
2001, \mnras, 328, 393

\bibitem[Aharonian(2001)]{aharonian01}
Aharonian,~F. 2001, in Proc.~27th Int. Cosmic Ray Conf. (Hamburg), 250

\bibitem[Aharonian et al.(2003)]{aharonian03}
Aharonian,~F., et al. 2003, \aap, 403, L1

\bibitem[Bai \& Lee(2001)]{bai01}
Bai,~J.~M., \& Lee,~M.~G. 2001, \apj, 549, L173

\bibitem[Beilicke et al.(2005)]{beilicke05}
Beilicke,~M., et al. 2005, in Proc.~29th Int. Cosmic Ray Conf.
(Pune), 4, 299

\bibitem[Biermann \& Strittmatter(1987)]{biermann87}
Biermann,~P.~L., \& Strittmatter,~P.~A. 1987, \apj, 322, 643

\bibitem[Biretta et al.(1991)]{biretta91}
Biretta,~J.~A., Stern,~C.~P., \& Harris,~D.~E. 1991, \aj, 101, 1632

\bibitem[Biretta et al.(1995)]{biretta95}
Biretta,~J.~A., Zhou,~F., \& Owen,~F.~N. 1995, \apj, 447, 582

\bibitem[Capetti et al.(1997)]{capetti97}
Capetti,~A., Macchetto,~F.~D., Sparks,~W.~B., \& Biretta,~J.~A.
1997, \aap, 317, 637

\bibitem[Carilli \& Barthel(1996)]{carilli96}
Carilli,~C.~L., \& Barthel,~P.~D. 1996, \aapr, 7, 1

\bibitem[Coppi \& Aharonian(1997)]{coppi97}
Coppi,~P.~S., \& Aharonian,~F.~A. 1997, \apj, 487, L9

\bibitem[Di~Matteo et al.(2003)]{dimatteo03}
Di~Matteo,~T., Allen,~S.~W., Fabian,~A.~C., Wilson,~A.~S.,
\& Young,~A.~J. 2003, \apj, 582, 133

\bibitem[Fleishman(2006)]{fleishman06}
Fleishman,~G.~D. 2006, \mnras, 365, L11

\bibitem[Gould \& Schr\'eder(1966)]{gould66}
Gould,~R.~J., \& Schr\'eder,~G. 1966, \prl, 16, 252

\bibitem[Heinz \& Begelman(1997)]{heinz97}
Heinz,~S., \& Begelman,~M.~C. 1997, \apj, 490, 653

\bibitem[Hofmann(2001)]{hofmann01}
Hofmann,~W. 2001, in Proc.~27th Int. CosmicvRay Conf.
(Hamburg), 2785

\bibitem[Honda \& Honda(2002)]{honda02}
Honda,~M., \& Honda,~Y.~S. 2002, \apj, 569, L39

\bibitem[Honda \& Honda(2004a)]{honda04a}
------. 2004a, \apj, 617, L37

\bibitem[Honda \& Honda(2005a)]{honda05a}
------. 2005a, \apj, 633, 733

\bibitem[Honda et al.(2000a)]{honda00a}
Honda,~M., Meyer-ter-Vehn,~J., \& Pukhov,~A.
2000a, Phys.~Plasmas, 7, 1302

\bibitem[Honda et al.(2000b)]{honda00b}
------. 2000b, \prl, 85, 2128

\bibitem[Honda \& Honda(2004b)]{honda04b}
Honda,~Y.~S., \& Honda,~M. 2004b, \apj, 613, L25

\bibitem[Honda \& Honda(2005b)]{honda05b}
------. 2005b, \mnras, 362, 833

\bibitem[Jokipii(1987)]{jokipii87}
Jokipii,~J.~R. 1987, \apj, 313, 842

\bibitem[Kardashev(1962)]{kardashev62}
Kardashev,~N.~S. 1962, Soviet Astron.--AJ, 6, 317

\bibitem[Kataoka et al.(2001)]{kataoka01}
Kataoka,~J., et al. 2001, \apj, 560, 659

\bibitem[Keel(1988)]{keel88}
Keel,~W.~C. 1988, \apj, 329, 532

\bibitem[Kneiske et al.(2004)]{kneiske04}
Kneiske,~T.~M., Bretz,~T., Mannheim,~K., \& Hartmann,~D.~H.
2004, \aap, 413, 807

\bibitem[Kneiske \& Mannheim(2005)]{kneiske05}
Kneiske,~T.~M., \& Mannheim,~K. 2005,
Proc.~29th Int. Cosmic-Ray Conf. (Pune), 4, 1

\bibitem[Le~Bohec et al.(2004)]{lebohec04}
Le~Bohec,~S., et al. 2004, \apj, 610, 156

\bibitem[Lobanov et al.(2003)]{lobanov03}
Lobanov,~A., Hardee,~P., \& Eilek,~J. 2003, \nar, 47, 629

\bibitem[Longair(1992)]{longair92}
Longair,~M.~S. 1992, High Energy Astrophysics, Vol.~1:
Particles, Photons and Their Detection
(Cambridge: Cambridge~Univ.~Press)

\bibitem[Longair(1994)]{longair94}
------. 1994, High Energy Astrophysics, Vol.~2:
Stars, the Galaxy and the Interstellar Medium
(Cambridge: Cambridge~Univ.~Press)

\bibitem[Marshall et al.(2002)]{marshall02}
Marshall,~H.~L., Miller,~B.~P., Davis,~D.~S., Perlman,~E.~S., Wise,~M.,
Canizares,~C.~R., \& Harris,~D.~E. 2002, \apj, 564, 683

\bibitem[McEnery et al.(2004)]{mcenery04}
McEnery,~J.~E., Moskalenko,~I.~V., \& Ormes,~J.~F. 2004,
in Cosmic Gamma-Ray Sources, ed. K.~S.~Cheng \& G.~E.~Romero
(Dordrecht; Kluwer), 361

\bibitem[Meisenheimer et al.(1996)]{meisenheimer96}
Meisenheimer,~K., R\"oser,~H.-J., \& Schl\"otelburg,~M. 1996, \aap, 307, 61

\bibitem[Montgomery \& Liu(1979)]{montgomery79}
Montgomery,~D., \& Liu,~C.~S. 1979, Phys.~Fluids, 22, 866

\bibitem[Nikishov(1962)]{nikishov62}
Nikishov,~A.~I. 1962, Sov.~Phys.--JETP, 14, 393

\bibitem[Ostrowski(2000)]{ostrowski00}
Ostrowski,~M. 2000, \mnras, 312, 579

\bibitem[Owen et al.(1990)]{owen90}
Owen,~F.~N., Eilek,~J.~A., \& Keel,~W.~C. 1990, \apj, 362, 449

\bibitem[Owen et al.(1989)]{owen89}
Owen,~F.~N., Hardee,~P.~E., \& Cornwell,~T.~J. 1989, \apj, 340, 698

\bibitem[P\'erez-Fournon et al.(1988)]{perez-fournon88}
P\'erez-Fournon,~I., Colina,~L., Gonz\'alez-Serrano,~J.~I., \& Biermann,~P.~L.
1988, \apj, 329, L81

\bibitem[Perlman et al.(2001)]{perlman01}
Perlman,~E.~S., Biretta,~J.~A., Sparks,~W.~B., Macchetto,~F.~D., \&
Leahy,~J.~P. 2001, \apj, 551, 206

\bibitem[Perlman et al.(1999)]{perlman99}
Perlman,~E.~S., Biretta,~J.~A., Zhou,~F., Sparks,~W.~B., \&
Macchetto,~F.~D. 1999, \aj, 117, 2185

\bibitem[Perlman \& Wilson(2005)]{perlman05}
Perlman,~E.~S., \& Wilson,~A.~S. 2005, \apj, 627, 140

\bibitem[Protheroe et al.(2003)]{protheroe03}
Protheroe,~R.~J., Donea,~A.-C., \& Reimer,~A. 2003, Astropart.~Phys., 19, 559

\bibitem[Protheroe \& Meyer(2000)]{protheroe00}
Protheroe,~R.~J., \& Meyer,~H. 2000, Phys.~Lett.~B, 493, 1

\bibitem[Reid et al.(1989)]{reid89}
Reid,~M.~J., Biretta,~J.~A., Junor,~W., Muxlow,~T.~W.~B, \& Spencer,~R.~E.
1989, \apj, 336, 112

\bibitem[Reimer et al.(2004)]{reimer04}
Reimer,~A., Protheroe,~R.~J., \& Donea,~A.~C. 2004, A\&A, 419, 89

\bibitem[Reimer et al.(2003)]{reimer03}
Reimer,~O., Pohl,~M., Sreekumar,~P., \& Mattox,~J.~R. 2003, \apj, 588, 155

\bibitem[Schlickeiser(2002)]{schlickeiser02}
Schlickeiser,~R. 2002, Cosmic Ray Astrophysics
(Berlin: Springer)

\bibitem[Silva et al.(2003)]{silva03}
Silva,~L.~O., Fonseca,~R.~A., Tonge,~J.~W., Dawson,~J.~M., Mori,~W.~B., \&
Medvedev,~M.~V. 2003, \apj, 596, L121

\bibitem[Sreekumar et al.(1998)]{sreekumar98}
Sreekumar,~P., et al. 1998, \apj, 494, 523

\bibitem[Stawarz et al.(2006a)]{stawarz06a}
Stawarz,~\L., Aharonian,~F., Kataoka,~J., Ostrowski,~M.,
Siemiginowska,~A., \& Sikora,~M. 2006a, \mnras, 370, 981

\bibitem[Stawarz et al.(2006b)]{stawarz06b}
Stawarz,~\L., Kneiske,~T.~M., \& Kataoka,~J. 2006b, \apj, 637, 693

\bibitem[Stawarz et al.(2003)]{stawarz03}
Stawarz,~\L., Sikora,~M., \& Ostrowski,~M. 2003, \apj, 597, 186

\bibitem[Stiavelli et al.(1991)]{stiavelli91}
Stiavelli,~M., M{\o}ller,~P., \& Zeilinger,~W.~W. 1991, \nat, 354, 132

\bibitem[Stiavelli et al.(1997)]{stiavelli97}
Stiavelli,~M., Peletier,~R.~F., \& Carollo,~C.~M. 1997, \mnras, 285, 181

\bibitem[Strong et al.(2004)]{strong04}
Strong,~A.~W., Moskalenko,~I.~V., \& Reimer,~O. 2004, \apj, 613, 956

\bibitem[Takahashi et al.(2004)]{takahashi04}
Takahashi,~T., et al. 2004, Proc.~SPIE, 5488, 549

\bibitem[Tonry(1991)]{tonry91}
Tonry,~J.~L. 1991, \apj, 373, L1

\bibitem[Toptygin \& Fleishman(1987)]{toptygin87}
Toptygin,~I.~N., \& Fleishman,~G.~D. 1987, \apss, 132, 213

\bibitem[Tsvetanov et al.(1998)]{tsvetanov98}
Tsvetanov,~Z.~I., Hartig,~G.~F., Ford,~H.~C., Dopita,~M.~A.,
Kriss,~G.~A., Pei,~Y.~C., Dressel,~L.~L., \& Harms,~R.~J.
1998, \apj, 493, L83

\bibitem[Vainio(1999)]{vainio99}
Vainio,~R. 1999, in Plasma Turbulence and Energetic Particles
in Astrophysics, ed. M.~Ostrowski \& R.~Schlickeiser
(Krak\'ow: Obs.~Astron.), 232

\bibitem[Wei et al.(2004)]{wei04}
Wei,~M.~S., et al. 2004, \pre, 70, 056412

\bibitem[Wilson \& Yang(2002)]{wilson02}
Wilson,~A.~S., \& Yang,~Y. 2002, \apj, 568, 133\\
\end{thebibliography}
\end{document}